\documentclass[10pt,twocolumn,twoside]{IEEEtran} 
\usepackage{amsmath,amsfonts}
\DeclareMathOperator*{\argmax}{arg\,max}

\usepackage{mathrsfs}
\usepackage{algorithm}
\usepackage{algpseudocode}
\usepackage{array}
\usepackage[font=footnotesize,labelfont=sf,textfont=sf]{subfig}
\usepackage{textcomp}
\usepackage{stfloats}
\usepackage{url}
\usepackage{verbatim}
\usepackage{graphicx}
\usepackage{cite}
\usepackage{acro}
\usepackage{xcolor}

\DeclareAcronym{wsn}{
    short=WSN,
    long=Wireless Sensor Network
}
\DeclareAcronym{iot}{
    short=IOT,
    long=Internet of Things
}
\DeclareAcronym{pdf}{
    short=PDF,
    long=probability density function
}
\DeclareAcronym{cdf}{
    short=CDF,
    long=cumulative distribution function
}
\DeclareAcronym{dof}{
    short=DOF,
    long=degrees of freedom
}
\DeclareAcronym{iid}{
    short=i.i.d.,
    long=independently and identically distributed
}
\DeclareAcronym{rdp}{
    short=RDP,
    long=Ramer-Douglas-Peucker
}

\DeclareAcronym{fisst}{
    short=FISST,
    long=finite set statistics
}
\DeclareAcronym{rfs}{
    short=RFS,
    long=Random Finite Set
}
\DeclareAcronym{mot}{
    short=MOT,
    long=multi object tracking
}
\DeclareAcronym{phd}{
    short=PHD,
    long=Probability Hypothesis Density
}
\DeclareAcronym{lmb}{
    short=LMB,
    long=labeled multi-Bernoulli
}
\DeclareAcronym{fov}{
    short=FOV,
    long=field of view
}
\DeclareAcronym{gm}{
    short=GM,
    long=Gaussian mixture
}
\DeclareAcronym{gc}{
    short=GC,
    long=Gaussian component
}
\DeclareAcronym{bc}{
    short=BC,
    long=Bernoulli component
}
\DeclareAcronym{ci}{
    short=CI,
    long=covariance intersection
}
\DeclareAcronym{emd}{
    short=EMD,
    long=exponential mixture
}
\DeclareAcronym{gci}{
    short=GCI,
    long=generalized covariance intersection
}
\DeclareAcronym{mil}{
    short=MIL,
    long=minimum information loss
}
\DeclareAcronym{ga}{
    short=GA,
    long=Geometric Average
}
\DeclareAcronym{aa}{
    short=AA,
    long=Arithmetic Average
}
\DeclareAcronym{md}{
    short=MD,
    long=Mahalanobis Distance
}
\DeclareAcronym{mle}{
    short=MLE,
    long=Maximum Likelihood Estimation
}
\DeclareAcronym{kf}{
    short=KF,
    long=Kalman filter
}
\DeclareAcronym{ekf}{
    short=EKF,
    long=extended Kalman filter
}
\DeclareAcronym{jpda}{
    short=JPDA,
    long=joint probabilistic data association
}
\DeclareAcronym{mht}{
    short=MHT,
    long=multi-hypothesis tracking
}
\DeclareAcronym{da}{
    short=DA,
    long=data association
}

\DeclareAcronym{kld}{
    short=KLD,
    long=Kullback-Leibler divergence
}

\DeclareAcronym{ospa}{
    short=OSPA,
    long=optimal sub-pattern assignment
}
\DeclareAcronym{gospa}{
    short=GOSPA,
    long=generalized optimal sub-pattern assignment
}
\DeclareAcronym{nees}{
    short=NEES,
    long=normalized estimation error squared
}
\DeclareAcronym{rms}{
    short=RMS,
    long=root mean square
}

\DeclareAcronym{cv}{
    short=CV,
    long=Constant Velocity
}

\DeclareAcronym{aoa}{
    short=AOA,
    long=angle of arrival
}
\DeclareAcronym{aod}{
    short=AOD,
    long=angle of departure
}
\DeclareAcronym{toa}{
    short=TOA,
    long=time of arrival
}
\DeclareAcronym{los}{
    short=LOS,
    long=line of sight
}

\DeclareAcronym{isac}{
    short=ISAC,
    long=integrated sensing and communication
}
\DeclareAcronym{ris}{
    short=RIS,
    long=reflective intelligent surface
}
\DeclareAcronym{bs}{
    short=BS,
    long=base station
}

\DeclareAcronym{nmpc}{
    short=NMPC,
    long=nonlinear model predictive control
}
\DeclareAcronym{qp}{
    short=QP,
    long=quadratic program
}
\DeclareAcronym{fdi}{
    short=FDI,
    long=False Data Injection
}
\DeclareAcronym{dm}{
    short=DM,
    long=Data Modification
}
\DeclareAcronym{glrt}{
    short=GLRT,
    long=Generalized Likelihood Ration Test
}
\DeclareAcronym{fd}{
    short=FD,
    long=Fault Diagnosis
}
\DeclareAcronym{mcs}{
    short=MCS,
    long=Monte Carlo Simulation
}
\DeclareAcronym{rng}{
    short=RNG,
    long=Random Number Generator
}
\DeclareAcronym{fpr}{
    short=FPR,
    long=false positive rate
}
\DeclareAcronym{tpr}{
    short=TPR,
    long=true positive rate
}
\DeclareAcronym{dos}{
    short=DOS,
    long=denial-of-service
}
\DeclareAcronym{cps}{
    short=CPS,
    long=cyber-physical system
}

\DeclareAcronym{gnss}{
    short=GNSS,
    long=global navigation satellite system,
}
\DeclareAcronym{gps}{
    short=GPS,
    long=global positioning system,
}
\DeclareAcronym{rf}{
    short=RF,
    long=radio frequency
}
\DeclareAcronym{uwb}{
    short=UWB,
    long=ultra-wide band
}
\DeclareAcronym{imu}{
    short=IMU,
    long=inertial measurement unit
}

\DeclareAcronym{uav}{
    short=UAV,
    long=unmanned aerial vehicle
}

\DeclareAcronym{mip}{
    short=MIP,
    long=mixed integer programming
}

\DeclareAcronym{ml}{
    short=ML,
    long=machine learning
}

\newtheorem{assum}{Assumption}

\begin{document}
\title{Multi-Hypotheses Navigation in Collaborative Localization subject to Cyber Attacks}

\author{Peter Iwer Hoedt Karstensen, Roberto Galeazzi \\ Technical University of Denmark, Control, Robotics and Embodied AI Group
}



\maketitle

\begin{abstract}
This paper addresses resilient collaborative localization in multi-agent systems exposed to spoofed radio frequency measurements. Each agent maintains multiple hypotheses of its own state and exchanges selected information with neighbors using covariance intersection. Geometric reductions based on distance tests and convex hull structure limit the number of hypotheses transmitted, controlling the spread of hypotheses through the network. The method enables agents to separate spoofed and truthful measurements and to recover consistent estimates once the correct hypothesis is identified. Numerical results demonstrate the ability of the approach to contain the effect of adversarial measurements, while also highlighting the impact of conservative fusion on detection speed. The framework provides a foundation for resilient multi-agent navigation and can be extended with coordinated hypothesis selection across the network.
\end{abstract}

\begin{IEEEkeywords}
Collaborative Localization, Cyber Attacks, Multi-Hypotheses Navigation, Covariance Intersection, Resilient Navigation, Multi-Agent System
\end{IEEEkeywords}

\section{Introduction}
Multi-agent systems, such as multi-robot systems, are projected to see widespread use. In recent years, multi-robot systems have attracted significant interest, demonstrating promising capabilities in applications such as search and rescue and surveillance \cite{scherer_multi-robot_2020, li_resilient_2024}.

Recently, research has largely pivoted toward making such systems resilient, with a particular emphasis on cyber-resiliency \cite{kargar_tasooji_secure_2022}. A multi-robot system is a cyber-physical system, and malicious actors may spoof or inject, modify, and delete information exchanged between robots. Cyber-resilient navigation is fundamental to achieving system-level cyber-resiliency.

Depending on the application, the infrastructure available to support a multi-robot system may be minimal, for example, consisting of only a few base stations or access points. In such cases, collaborative localization becomes especially important. If a malicious actor gains access to the system, it may be subjected to spoofing or other cyber-attacks. Cyber-resiliency refers to the ability of a system to withstand unforeseen events and recover, either by regaining its original performance or by maintaining a degraded but functional level of performance \cite{tzavara_tracing_2024}.

A multi-robot system is typically deployed to achieve a mission-oriented goal, such as spatio-temporal security surveillance. This goal is jeopardized when one or more robots are subjected to cyber-attacks that distort their pose estimates, causing them to believe they are at a certain location while they are, in reality, elsewhere. Furthermore, in line with the concept of resiliency, the system cannot anticipate when or where an attack may occur, or how extensive it might be. For the system to function, at least one reliable subset of measurements must be available from which accurate pose information can be inferred. 

To address this, the system must track multiple hypotheses of its state. In practice, most measurements may support one hypothesis, which could be the result of a spoofing attack, while only a minority of measurements correspond to the true state. By maintaining multiple hypotheses, robots can recover the true state and thereby restore system performance.

This paper extends the work in \cite{karstensen_multi-hypotheses_2025}. That prior work considered the scenario where any \ac{rf}-based measurement source, such as a \ac{gnss} receiver or \ac{rf} anchors, could be spoofed. This paper investigates the same spoofing scenario but extends it to a multi-agent system and in a \ac{gnss}-denied environment. The main contribution is the extension of the method proposed in \cite{karstensen_multi-hypotheses_2025}, where spoofed sources are grouped, to the domain of collaborative localization. The following assumptions in addition to those listed in \cite{karstensen_multi-hypotheses_2025} are used in the development of the proposed method.

\begin{assum}
\label{assum:attack_dir}
    The malicious entity may only make a joint attack in two distinct directions such that spoofed location together with the true location form the vertices of a convex hull
\end{assum}
\begin{assum}
\label{assum:sources}
    The malicious entity is more likely to attack the anchors due to their known location, for which reason the agents as an initial guess remove anchors from a selected hypothesis
\end{assum}

Robots and agents are used interchangeably in this paper.

\section{Related Work}

Collaborative localization is a paradigm in which individual robots in a network perform relative measurements to update their estimates and subsequently share their posteriors, fusing them to achieve more precise localization. Fundamentally, this tightly couples the robots, making their estimation processes dependent on one another. Ignoring this dependence often results in overconfident estimates. Such overconfidence leads to inconsistent estimates \cite{ wang_fault_2021, luft_recursive_2018, chang_resilient_2022}.  

To address this issue, numerous variants of the \ac{ekf} for collaborative localization have been proposed, each aiming to account for inter-robot dependence, often under additional constraints such as communication outages. These works typically compare the trade-offs between consistency and computational complexity. For example, the authors of \cite{zu_cooperative_2019} propose a fusion method that mimics \ac{ci} by computing covariance matrices that upper bound the true covariance. In \cite{luft_recursive_2018}, each robot updates a decomposed joint covariance matrix relative to its neighbors, intended to capture the dependence, however, the authors note that consistency is not guaranteed. A three-step update method is presented in \cite{chang_resilient_2022}, where robots maintain their own states and track their neighbors, sharing posteriors that are fused via \ac{ci}. In \cite{wang_fault_2021} propose a covariance union collaborative localization scheme, able to handle spurious inter-robot measurements, however, with the downside of being overly conservative. The authors of \cite{yan_robust_2024} consider inter-robot measurements that experience measurement outliers. They model the measurement noise as a Student's $t$-distribution and though variational Bayesian iteration derive a scalar, that scales the measurement covariance matrix, lowering the contribution of distant measurement in terms of the innovation signal. 

Beyond estimator consistency, the threat of malicious interference has been a major research focus. Historically, attacks on navigation systems have concentrated on \ac{gnss} spoofing due to the relatively low effort required by an attacker \cite{xu_sok_2023, venturino_adaptive_2025}. For instance, \cite{venturino_adaptive_2024} proposes a multi-antenna \ac{gnss} receiver that uses \ac{aoa} information to determine whether a satellite signal is spoofed. In \cite{yoon_towards_2019}, a $\chi^2$-detector is employed to detect spoofing by cross-validating \ac{gnss} with \ac{imu} data. A multi-robot scenario is considered in \cite{michieletto_robust_2023}, where robots operate in a region containing a spoofed subregion. While robots primarily navigate using \ac{imu} and \ac{gnss}, they also perform relative ranging, which enables hypothesis testing to verify the integrity of \ac{gnss} measurements. By maintaining additional state variables, the robots switch adaptively between \ac{gnss} and relative measurements.  

In contrast, limited work has focused on spoofing of communication infrastructure that provides positional measurements. In \cite{he_robust_2024}, an M-estimator is proposed to mitigate locally biased measurements at each agent, after which posteriors are fused using \ac{ci}. The work in \cite{salimpour_exploiting_2023} considers a system in which one unknown node produces abnormal measurements; the faulty node is isolated by computing $n-1$ estimates and identifying the configuration with the highest variance. Similarly, \cite{vijay_range-based_2025} investigates spoofing of relative measurements, which may lead to safety constraint violations. The authors formulate a global optimization problem to estimate the attacks, approximated via distributed optimization. Related but not identical, network-wide fault detection is studied in \cite{al_hage_multi-sensor_2017}, where fault diagnosis is used to isolate faults in robots’ odometry and sensing devices. Event-triggered collaborative localization under adversarial conditions is explored in \cite{kargar_tasooji_secure_2022}, where robots share measurements and priors that may be manipulated by an attacker. The decision to accept neighbor information is based on a threshold applied to the innovation norm, though the approach introduces several tuning parameters without guidance for selecting them. The method is extended in \cite{tasooji_distributed_2025} to handle subregions in which cyber-attacks may target either communication links or sensors. The authors of Fault-tolerant collaborative localization is also considered in \cite{qu_fault-tolerant_2017}, where a team of robots relies on both collaborative localization and GNSS. Two filters are maintained and fused using covariance intersection, with weights determined by the normalized innovation error squared so that inconsistent measurements contribute less. The authors assume that GNSS and collaborative localization cannot be faulty simultaneously, and a chi-squared test is used to determine which source becomes erroneous.

A common assumption across the literature is that at least 50\% of the measurement sources are uncompromised \cite{vijay_range-based_2025}. This ensures a sufficient number of truthful measurements to maintain localization. When multiple hypotheses are tracked, robots can often recover accurate localization despite attacks. Such ideas date back to early work in simultaneous localization and mapping \cite{arras_feature-based_2002}. In \ac{fd}, a similar approach involves maintaining a bank of filters, each using a subset of measurements \cite{gertler_survey_1988}. This approach lost popularity due to computational overhead and has been left untouched since. Recently, this approach was applied to navigation in \cite{jurado_residual-based_2020} considering a single fault and two faults in \cite{gipson_resilience_2022}, and extended to consider collaborative localization in \cite{gipson_framework_2022}. The method maintains two separate bank of filters, one utilizing proprioceptive  sensors only and the second utilizing collaborative localization only, both active at the same time and constructing filter utilizing $I-1$ and $I-2$ measurements, where $I$ is the number of measurements. The authors propose using the interleaving update algorithm of \cite{bahr_consistent_2009} which in itself constructs $2^I$ filters since it needs to track the contribution of each agent to the specific filter. Also, it requires. 

Recently, it has been demonstrated how an adversary can spoof the positioning capabilities within a cellular network \cite{li_ris-aided_2025}.

\section{Multi Agent Network}
This paper considers a network of mobile agents connected through some \ac{rf} network. The agents form a graph $\mathcal{G} = (\mathcal{P},\mathcal{E})$, where the agents are the vertices $\mathcal{P}$, which encodes their poses $\mathcal{P}\ni \mathbf{q}_k^{(i)} = \begin{bmatrix} \left(\mathbf{p}_k^{(i)}\right)^\mathrm{T} & \theta_k^{(i)}\end{bmatrix}^\mathrm{T}$, where $\mathbf{p}_k^{(i)} = \begin{bmatrix}x_k^{(i)} & y_k^{(i)}\end{bmatrix}^\mathrm{T}$. A subset of the vertices in the graph are static representing anchors, $\mathcal{P}_\mathrm{RF} \in \mathcal{P}$, whose position and orientation are known to all other agents in the network. The neighborhood of agent $i$ is denoted as $\mathcal{P}_k^{(i)}=\left\{j\mid ||\mathbf{p}_k^{(i)}-\mathbf{p}_k^{(j)}||_2<\rho\right\}$, where $\rho$ is the communication range of the agents and $||\cdot||_2$ is the 2-norm. The $i$th agent is referred to as $a^{(i)}$.

The state of the agents is modeled in discrete time as
\begin{equation}
    \label{eq:motion}
    \mathbf{x}_{k}^{(i)} = \mathbf{f}\left(\mathbf{x}_{k-1}^{(i)},\mathbf{u}_k^{(i)}\right) + \mathbf{w}_{\mathrm{d},k}^{(i)},
\end{equation}
where $\mathbf{x}_k^{(i)}$ includes the agent's pose, $\mathbf{u}_k^{(i)}$ are the inputs to the system such as \ac{imu} measurements, and $\mathbf{w}_{\mathrm{d},k}^{(i)}$ is white Gaussian noise with covariance matrix $\mathbf{Q}_k^{(i)}$ independent of any other noise sources.

Each agent track its neighbors, $\mathcal{P}_k^{(i)}$, via the following discrete model 
\begin{equation}
    \mathbf{x}_k^{(i,j)} = \mathbf{g}_k\left(\mathbf{x}_k^{(i,j)},\mathbf{u}_k^{(j)}\right) +\mathbf{w}_{\mathrm{d},k}^{(i,j)},
\end{equation}
where $\mathbf{x}_k^{(i,j)}$ includes the pose of the neighbour $j$, $\mathbf{u}_k^{(j)}$ is the input to the neighbour $j$, such as \ac{imu} measurements, transmitted to agent $i$ and $\mathbf{w}_{\mathrm{d},k}^{(j)}$ is white Gaussian noise with covariance matrix $\mathbf{Q}_k^{(i,j)}$, representing tracking errors. This is distinct from $\mathbf{w}_{\mathrm{d},k}^{(j)}$, since the motion model in \eqref{eq:motion} may differ from the tracking model. To reduce the bandwidth, pre-integrated \ac{imu} measurements should be transmitted \cite{forster_-manifold_2017}.

When agent $i$ is within communication range of agent $j$, it performs measurements as follows \cite{karstensen_multi-hypotheses_2025}
\begin{equation}
    \label{eq:meas}
    \mathbf{z}_k^{(i,j)} = \mathbf{h}\left(\mathbf{q}_k^{(i)}, \mathbf{q}_k^{(j)},\boldsymbol{\epsilon}_k^{(i,j)}\right) + \mathbf{w}_{\mathrm{m},k}^{(i,j)},
\end{equation}
where
\begin{align}
    &\mathbf{z}_k^{(i,j)} =
    \begin{bmatrix}
        r_k^{(i,j)} & \theta_{\mathrm{AOA},k}^{(i,j)} & \theta_{\mathrm{AOD},k}^{(i,j)}
    \end{bmatrix}^\mathrm{T}, \\
    &\mathbf{h}\left(\mathbf{q}_k^{(i)},\mathbf{q}_k^{(j)},\boldsymbol{\epsilon}_k^{(i,j)}\right)= \\
    &
    \begin{bmatrix}
        ||\mathbf{p}_k^{(j)} - \mathbf{p}_k^{(i)} + \boldsymbol{\epsilon}_k|| \\ \pi +\mathrm{tan}^{-1}\left(y_k^{(j)} - y_k^{(i)} + \epsilon_{y,k}, x_k^{(j)} - x_k^{(i)} + \epsilon_{x,k}\right) -\theta_k^{(i)} \\
        \mathrm{tan}^{-1}\left(y_k^{(j)} - y_k^{(i)} + \epsilon_{y,k}, x_k^{(j)} - x_k^{(i)} +\epsilon_{x,k}\right) -\theta_k^{(j)}
    \end{bmatrix}.
\end{align}
$r^{(i,j)}$ is the inter-agent distance. $\theta_\mathrm{AOA}^{(i,j)}$ is the \ac{aoa} at agent $i$. $\theta_\mathrm{AOD}^{(i,j)}$ is the \ac{aod} at agent $j$. $\mathbf{w}_{\mathrm{RF},k}^{(i,j)}$ is white Gaussian noise with zero expected value and covariance matrix $\mathbf{R}_{\mathrm{RF}}$. $\boldsymbol{\epsilon}_k = \begin{bmatrix} \epsilon_{x,k} & \epsilon_{y,k}\end{bmatrix}$ is some unknown time-varying adversarial signal. The measurements are associated with an identifier $o_k^{(i)}$.

In some regions, agents may localize relative to an \ac{rf} anchor. The same measurement model as in \eqref{eq:meas} applies, except that the anchor’s position and orientation are assumed known.

Each agent tracks nominal trajectores $\mathbf{t}_\mathrm{n}^{(i)}$, which are sequences of time-indexed poses.

\subsection{Multi Hypotheses Navigation}

In \cite{karstensen_multi-hypotheses_2025}, a method was presented for scenarios where any measurement source, in any number, may be spoofed. This lead to a multi-hypothesis formulation, where each hypothesis utilizes a subset of the measurements. The hypotheses are compactly expressed as
\begin{equation}
    \label{eq:hypo}
    \mathcal{H}_{k}^{(i)} = \left\{h^{(i,\iota)}_k\right\}_{\iota=1}^{|\mathcal{H}_{k}^{(i)}|}, \quad h^{(i,\iota)}_k = \left(\left(\boldsymbol{\mu}_{(\cdot)}^{(i,\iota)},\mathbf{P}_{(\cdot)}^{(i,\iota)}\right),\mathcal{O}_k^{(i,\iota)}\right),
\end{equation}
where $\boldsymbol{\mu}_{(\cdot)}^{(i,\iota)}$ and $\mathbf{P}_{(\cdot)}^{(i,\iota)}$ denote the mean  and covariance matrix, respectively, of the prior, predicted prior, or posterior of the Gaussian hypothesis indexed with $\iota$. $\mathcal{O}_k^{(i,\iota)}$ is the subset of utilized measurements by hypothesis $h_k^{(\iota)}$. If a hypothesis considers all measurement sources, $\mathcal{O}_k^{(i,\iota)}=\mathcal{P}_k^{(i)}$. 

The method in \cite{karstensen_multi-hypotheses_2025} removes and generates hypotheses checking the consistency of the measurements with the hypotheses. It does so by counting the number of measurement outliers over a window $W$, and when this count exceeds an allowed number of outliers, then the hypothesis is rejected. When $h^{(i,\iota)}$ is rejected, $|\mathcal{O}_k^{(i,\iota)}|-1$ hypotheses are generated inheriting the Gaussian parameters where the covariance matrix is inflated with $\alpha_\mathrm{d} > 1$ and for each a measurement source in $\mathcal{O}^{(i,\iota)}_k$ are removed. Reduction methods are established such that no replicates of hypotheses in terms of $\mathcal{O}_k^{(i,\iota)}$. The method constructs a validation region $\mathcal{E}_{\mathbf{R}_\mathrm{RF,\mathrm{q}}}^{\gamma_{\alpha_\chi}}(\mathbf{z}_{k,\mathrm{q}}^{(i)})$ around each measurement and outliers for each predicted measurement $\hat{\mathbf{z}}_{k|k-1,\mathrm{q}}^{(\iota,i)} \not\in \mathcal{E}_{\mathbf{R}_\mathrm{RF,\mathrm{q}}}^{\gamma_{\alpha_\chi}}(\mathbf{z}_{k,\mathrm{q}}^{(i)})$ are counted over a window $W$, where $q$ is the $q$'th component of the vector or diagonal of a matrix. The binary variable encoding whether the measurement is an outlier is a Bernoulli random variable whose probability can be computed through the motion and measurement model. The outlier probability of measurement $j$'s $q$ component with respect to hypothesis $\iota$ of agent $i$ is denoted as $P_{\mathrm{out},k,\mathrm{q}}^{(i,\iota,j)}$. The sum of these Bernoulli random variables will follow a Poisson-binomial distribution. The maximum allowed number of outliers can then be determined when selecting a percentile $\beta$ as follows
\begin{equation}
    o_{\beta,q}^{(\iota,i)} = F_{\mathrm{PB}}^{-1}\left(\beta; W, \left\{P_{\mathrm{out},k,q}^{(\iota,i)}\right\}_{k=1}^{W}\right).
\end{equation}

An operational hypothesis $h_k^{(i,\mathrm{op})}$ is maintained, using all measurements.

\subsection{Hypotheses Filtering}

The agents employ an \ac{ekf} to estimate their own state along with the states of their neighbors. 

Each agent forms a complete state vector, $\mathbf{s}_k^{(i,\iota)}$, similar to \ac{ekf}-SLAM, as
\begin{equation}
    \mathbf{s}_k^{(i,\iota)} = 
    \begin{bmatrix} \boldsymbol{\mu}_k^{(i,\iota)} &    \mathrm{Vec}\left[\boldsymbol{\mu}_k^{(i,\iota,j)}\right]_{j\in\mathcal{O}_k^{(i,\iota)}}^\mathrm{T} 
    \end{bmatrix}^\mathrm{T},
\end{equation}
where $\mathrm{Vec}[\cdot]$ concatenates the neighbor states. The associated covariance matrix is denoted $\boldsymbol{\Sigma}_k^{(i,\iota)}$.

State prediction is carried out using \ac{imu} measurements from the agent itself and transmitted measurements from neighbors. The update step is performed using the measurement model in \eqref{eq:meas}, relative to neighbors and anchors. Agents then share portions of their posteriors with neighbors and locally fuse their own posteriors with those received. Details of this procedure follow in the next section.

\section{Resilient Collaborative Localization}

\begin{figure}
    \centering
    \includegraphics[width=\linewidth]{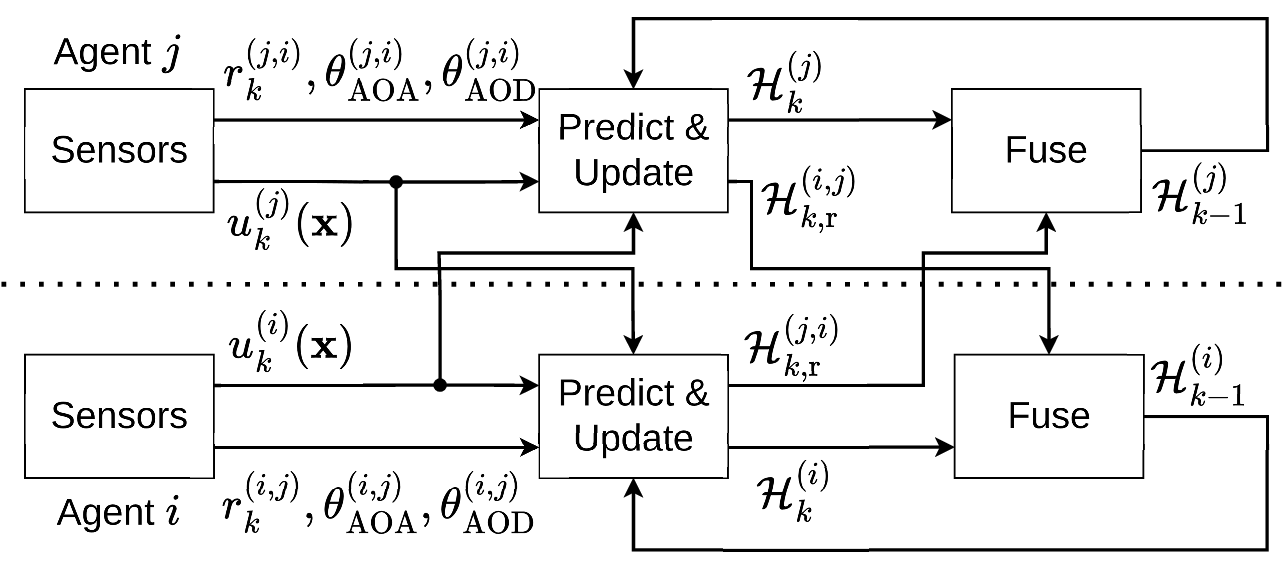}
    \caption{The information exchanged by the agents are the \ac{imu} measurements and the posteriors.}
    \label{fig:MHET_MAS}
\end{figure}

Once agent $i$ performs a relative measurement to neighbor $j$, after the Kalman filter update step the agents exchange posteriors, which are subsequently fused using \ac{ci}. This is visualized in Fig. \ref{fig:MHET_MAS}.

Directly applying the framework in \cite{karstensen_multi-hypotheses_2025} framework to a multi-agent system would limit a agent with only other agents as neighbors to a single hypothesis. This is due to the  reduction methods that do not allow replicates in term of $\mathcal{O}_k^{(i,\iota)}$. When an anchor measurement or more are spoofed somewhere in the network, a agent not in range of these anchors will eventually be presented with multiple hypotheses by one of its neighbors. This will have to prompt it to create additional hypotheses with a set of measurement support which already exists in its set of hypotheses. This is a way of spreading the hypotheses across the network, and that the neighbor is non-adversarial but that it simply also is spreading the hypotheses across the network. To still maintain the reduction methods, the hypotheses need to have an additional identifier to distinguish in those cases.

\subsection{Tagging of Hypotheses}

This paper extends the hypotheses of agent $i$ such that it now contains a tag $t$ as follows
\begin{equation}
    \label{eq:agent_hypo}
    \begin{split}
    \mathcal{H}_k^{(i)}&=\left\{h_{k}^{(i,\iota,t(\iota))}\right\}_{\iota=0}^{|\mathcal{H}_k^{(i)}|}, \\
    h^{(i,\iota)}_k &= \left(\left(\mathbf{s}_{(\cdot)}^{(i,\iota, t(\iota))},\mathbf{P}_{(\cdot)}^{(i,\iota,t(\iota)}\right),\mathcal{O}_k^{(i,\iota,t(\iota))}\right),
    \end{split}
\end{equation}
where $t(\iota)$ denotes the tag of hypothesis $\iota$. During reduction, only hypotheses with identical tags are considered. At time step $k$, the number of tags is given by  
\begin{equation}
    T_k^{(i)} = \left|\left\{t \mid t(\iota)\in \mathbb{N}, h_k^{(i,\iota,t(\iota))}\in H_k^{(i)}\right\}\right|.
\end{equation}
Initially, all hypotheses share one tag until agent $i$ receives multiple consecutive hypotheses from a neighbor. The tagging logic is described in Section \ref{sec:utilizing}.  

Each tag $t$ has an associated operational hypothesis $h_k^{(i,t,\mathrm{op})}$. Its measurement support is defined as  
\begin{equation}
    \mathcal{O}_k^{(t,\mathrm{op})} = \left\{o^{(j)} \mid j\in\mathcal{P}^{(i)}\cap\mathcal{O}_k^{(i,\iota,t(\iota))}, \;\iota=0,\dots,|\mathcal{H}_k|\right\}.
\end{equation}
The tag and how the hypotheses and operational hypotheses are related is illustrated in Table \ref{tab:hypo-table}.
\begin{table}[]
    \centering
    \begin{tabular}{l|l|l|l}
    $t$             & $0$ & $1$ & $2$\\ \hline
    $\mathrm{op}$ &  $h^{i,0,\mathrm{op})}$ & $h^{i,1,\mathrm{op})}$ & $h^{i,2,\mathrm{op})}$  \\ \hline
    $\mathcal{H}$ &  $h^{(i,0,0)}$ & $h^{(i,3,1)}$ & $h^{(i,5,2)}$  \\
                  &  $h^{(i,1,0)}$ & $h^{(i,4,1)}$ & $h^{(i,6,2)}$ \\
                  &  $h^{(i,2,0)}$ &               & $h^{(i,7,2)}$ 
    \end{tabular}
    \caption{Tag of hypotheses shown as a hypotheses table. Each tag has an operational hypotheses and a set of hypotheses.}
    \label{tab:hypo-table}
\end{table}

\subsection{Transmitting Information}
\label{sec:transmitting}

In previous work on collaborative localization using \ac{ci}, agent $i$ shares its complete state vector $\mathbf{s}_k^{(i)}$ with agent $j$ \cite{chang_resilient_2022}, which contains information about agent $i$s neighbors beyond agent $j$, and potentially have a common other neighbor. This allows agent $i$ to potentially skew the estimates of agent $j$ and a specific common neighbor. Therefore, the information sharing is limited to the state contents related to agent $i$ and $j$. 

Hypotheses supported by measurements relative to neighbor $j$ are selected via the index set  
\begin{equation}
    I_k^{(i,j)} = \left\{\iota \mid j\in\mathcal{O}_k^{(i,\iota,t(\iota))},\; \iota=1,\dots,|\mathcal{H}_k^{(i)}| \right\}.
\end{equation}
Agent $i$ constructs a matrix $\mathbf{T}_{\mathrm{s}}^{(i,\iota,j)}$ that extracts only the entries relating to $i$ and $j$ in $\mathbf{s}_k^{(i,\iota)}$, i.e.,  
\begin{equation}
    \begin{split}
        \mathbf{s}_{k,\mathrm{s}}^{(i,\iota,j)} &= \mathbf{T}_{\mathrm{s}}^{(i,\iota,j)}\mathbf{s}_k^{(i,\iota)}, \\
        \boldsymbol{\Sigma}_{k,\mathrm{s}}^{(i,\iota,j)} &= \mathbf{T}_{\mathrm{s}}^{(i,\iota,j)}\boldsymbol{\Sigma}_k^{(i,\iota,j)}\left(\mathbf{T}_{\mathrm{s}}^{(i,\iota,j)}\right)^\mathrm{T},
    \end{split} \quad \text{for} \;\iota \in I_k^{(i,j)}.
\end{equation}
Although the dimension of $\mathbf{s}_k^{(i,\iota)}$ may vary, $\mathbf{s}_k^{(i,\iota,j)}$ has consistent dimension across all $\iota$.  

\begin{figure}[t]
    \centering
    \includegraphics[width=1\linewidth]{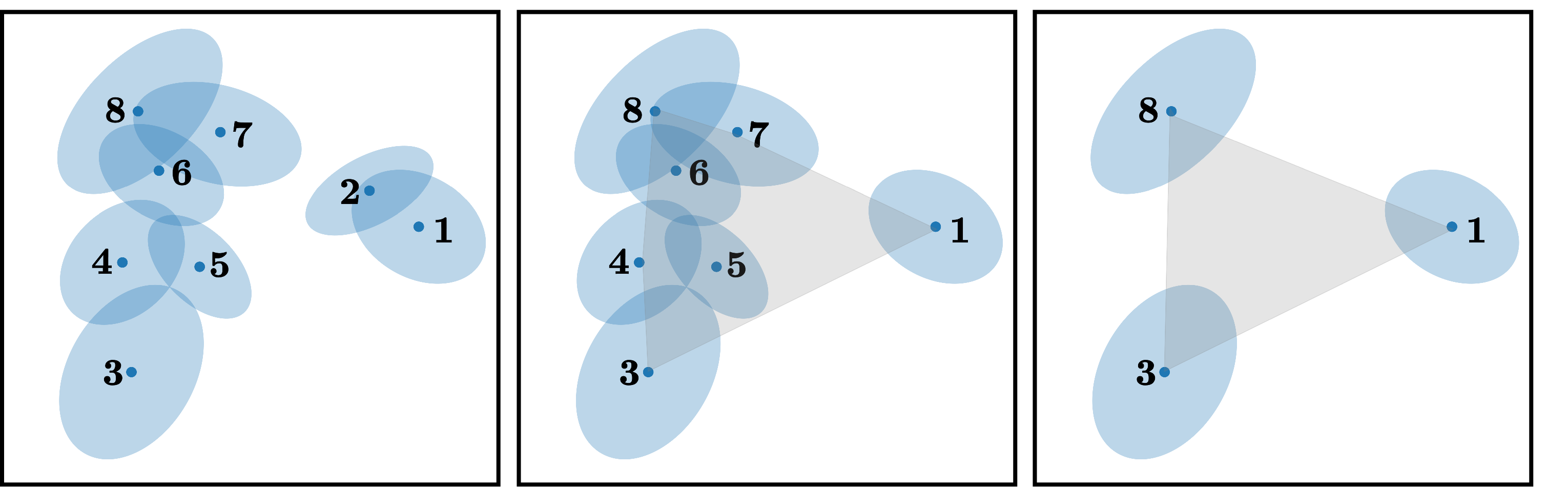}
    \caption{The procedure of determining the set of hypotheses transmitted to the neighbours. The middle box shows that first through $\mathbf{B}_k^{(i,j)}$, similar hypotheses are found such as hypotheses numbered $1$ and $2$ are considered the same, where $1$ is retained due to its distance to the remaining hypotheses higher is higher than $2$. The convex hull has hypotheses $1$, $3$, $4$, $7$ and $8$ as vertices. The right box shows how the hypotheses not constituting a vertex in the convex hull are not considered. The \ac{rdp} along with reductions on $\mathbf{B}_k^{(i,j)}$ remove vertices $4$ and $7$, falling inside the strip defined by $\varepsilon$.}
    \label{fig:transmitting_hypo}
\end{figure}

An agent continuously splits hypotheses. Waiting for the diagnosis to conclude before transmitting a set of hypotheses would delay diagnosis across the network. However, transmitting all hypotheses $\iota \in I_k^{(i,j)}$ increases fusion complexity and communication overhead. Thus, hypotheses must be reduced before transmission. Building on Assumption \ref{assum:attack_dir}, the most distant hypotheses are selected. Since malicious entities bias poses toward extremes, the hypotheses representing the true state and those representing spoofed states eventually form vertices of a convex hull in terms of the positional estimates at each agents. The procedure will be detailed in the following and is illustrated in Fig. \ref{fig:transmitting_hypo}.

To identify the vertices of the convex hull, a \ac{md} matrix $\mathbf{D}_{k,\mathrm{s}}^{(i,j)}$ is first formed across all hypotheses at agent $i$, computed on the marginalized poses of the agent $i$. Denote $\mathbf{s}_{k,\mathrm{s},\mathbf{q}}^{(i,j,\iota)}$ as the marginalized states at agent $i$ containing agent $i$s pose, with associated covariance matrix $\boldsymbol{\Sigma}_{k,\mathrm{s},\mathbf{q}}^{(i,j,\nu)}$. The entries of the \ac{md} matrix is 
\begin{equation}
    \mathbf{D}_{k,(\iota,\nu)}^{(i,j)} = d^{(\iota,\nu)}, \quad \iota,\nu \in I_k^{(i,j)},
\end{equation}
where $d^{(\iota,\nu)} = \left(\mathbf{e}_k^{(i,j,\iota,\nu)}\right)^\mathrm{T}\left(\boldsymbol{\Sigma}_{k,\mathrm{s},\mathbf{q}}^{(i,j,\iota)} + \boldsymbol{\Sigma}_{k,\mathrm{s},\mathbf{q}}^{(i,j,\nu)}\right)^{-1}\mathbf{e}_k^{(i,j,\iota,\nu)}$ and $\mathbf{e}_k^{(i,j,\iota,\nu)} = \mathbf{s}_{k,\mathrm{s},\mathbf{q}}^{(i,j,\iota)} - \mathbf{s}_{k,\mathrm{s},\mathbf{q}}^{(i,j,\nu)}$.  

Assuming $d^{(\iota,\nu)}$ follows a $\chi^2$-distribution, the entries are compared against the inverse \ac{cdf} at percentile $\alpha_\mathrm{T}$ and with three degrees of freedom. This yields a binary matrix $\mathbf{B}_k^{(i,j)}$. Denote $J_\iota = \left\{\nu\mid\mathbf{b}_\nu =\mathbf{b}_\iota\right\}\cup\{\iota\}$, where $\mathbf{b}_\nu$ is the $\nu$ row of $\mathbf{B}_k^{(i,j)}$. Hypotheses with identical rows, in which case $|J_\iota| > 1$, are clustered. Within each cluster the hypothesis with maximum distance to any other hypothesis is retained. Denote $\mathbf{D}_{k}^{(i,j,J_\nu)} = \mathbf{D}_{k,(J_\nu,J_\nu)}^{(i,j)}$ as the matrix which only contains rows and columns according to the index set $J_\nu$, the hypotheses with the maximum distance is stated as $\iota_{J_\nu} = \argmax_{l}\max_m \mathbf{D}_{k,l,m}^{(i,j,J_\nu)}$. This reduces the index set $I_k^{(i,j)}$, such that the index set $I_k^{(i,j)}\setminus\left\{\iota_{J_\nu}\mid |J_\nu| > 1\right\}$ reduces $\mathbf{D}_{k}^{(i,j)}$ and $\mathbf{B}_k^{(i,j)}$ further.

The convex hull of the remaining hypotheses is then determined via Quickhull \cite{barber_quickhull_1996} using only the position of agent $i$ in $\mathbf{s}_{k,\mathrm{s},\mathbf{q}}^{(i,j,\iota)}$ denoted as $\mathbf{s}_{k,\mathrm{s},\mathbf{p}}^{(i,j,\iota)}$. The \ac{rdp} algorithm \cite{ramer_iterative_1972} reduces the vertex set, removing a vertex that falls inside a strip between two other vertices. The width of the strip $\varepsilon$ is user-defined. This further reduces $I_k^{(i,j)}$. Again, hypotheses with identical rows in $\mathbf{B}_k^{(i,j)}$ following the procedure described in the previous paragraph. The procedures are summarized in Algorithm \ref{alg:transmitHypotheses}.

\begin{algorithm}[H]
    \caption{Determine transmitting hypotheses}
    \label{alg:transmitHypotheses}
    \begin{algorithmic}
        \Require $\mathbf{D}_k^{(i,j)}, \mathbf{B}_k^{(i,j)}, I_{k}^{(i,j)}, \;\mathbf{s}_{k,\mathrm{s},\mathbf{q}}^{(i,j,\iota)}$
        \Ensure $\mathcal{H}_{k,\mathrm{s}}^{(i,j)}$
        \State $\mathbf{D},\;\mathbf{B},\;I \gets$ \Call{Cluster}{$\mathbf{D}_k^{(i,j)}, \mathbf{B}_k^{(i,j)}, I_{k}^{(i,j)}$}
        \State $\mathbf{s}^{(I)} \gets \mathbf{s}_{k,\mathrm{s},\mathbf{p},(I,I)}^{(i,j,\iota)}$
        \State $\mathbf{s}^{(I_\mathrm{CH})}, \;I_\mathrm{CH} \gets$ QuickHull$\left(\mathbf{s}^{(I)}\right)$
        \State $\mathbf{s}^{(I_\mathrm{RDP})}, \;I_\mathrm{RDP} \gets$RDP($\mathbf{s}^{(I_\mathrm{CH})}, \; I_\mathrm{CH}$)
        \State $\mathbf{B} \gets \mathbf{B}_{(I_\mathrm{RDP},I_\mathrm{RDP})}, \; \mathbf{D}\gets \mathbf{D}_{(I_\mathrm{RDP},I_\mathrm{RDP})}$
        \State $\mathbf{D},\;\mathbf{B},\;I \gets$ \Call{Cluster}{$\mathbf{D}, \mathbf{B}, I_\mathrm{RDP}$}
        \State \textbf{return} hypotheses indexed by $I$ 
    \Procedure{Cluster}{$\mathbf{D}$, $\mathbf{B}$, $I$}
        \State $J_\nu \gets \{\iota \mid \mathbf{b}_\iota = \mathbf{b}_\nu\} \cup \{\nu\}$ \Comment{With $\mathbf{b}_\nu$ the $\nu$ row of $\mathbf{B}$}
        \State $\mathbf{D}^{(J_\nu)} \gets \mathbf{D}_{(J_\nu,J_\nu)}$
        \State $\iota_{J_\nu} = \argmax_l\max_m \mathbf{D}^{(J_\nu)}$
        \State $I \gets I \setminus \{\iota_{J_\nu} \mid |J_\nu| > 1\}$
        \State $B \gets B_{I,I}, \quad D \gets D_{I,I}$
        \State \textbf{return} $\mathbf{D}, \;\mathbf{B}, \; I$
    \EndProcedure
    \end{algorithmic}
\end{algorithm}

\subsection{Utilizing Received Information}
\label{sec:utilizing}

Each agent receives a set of hypotheses from its neighbors and needs to associate these to its own hypotheses. This is done using the operational hypotheses of each tag $t$, denoted $h_k^{(i,t,\mathrm{op})}$, and the Hungarian algorithm.  

At time $k$, agent $i$ has $T_k^{(i)}$ tags and receives Gaussian parameters from neighbor $j$ 
\begin{equation}
    \mathcal{H}_{k,\mathrm{r}}^{(i,j)} = \left\{\left(\boldsymbol{\mu}_{k,\mathrm{r}}^{(i,\iota,j)},\boldsymbol{\Sigma}_{k,\mathrm{r}}^{(i,\iota,j)}\right)\right\}_{\iota=0}^{|H_{k,\mathrm{r}}^{(i,j)}|}.
\end{equation}
Agent $a^{(i)}$ computes a cost matrix whose entries are \ac{md}s between marginalized pose states:
\begin{equation}
    \label{eq:match_cost}
    \mathbf{C}_{k,\mathrm{r},t,\iota}^{(i,j)} = d_{k,\mathrm{r}}^{(i,t,j,\iota)},
\end{equation}
where $d_{k,\mathrm{r}}^{(i,t,j,\iota)} = \left(\mathbf{e}_k^{(i,j,t,\iota)}\right)^\mathrm{T}\left(\boldsymbol{\Sigma}_k^{(i,j,t,\iota)} \right)^{-1}\mathbf{e}_k^{(i,j,t,\iota)}$, $\mathbf{e}_k^{(i,j,t,\iota)} = \mathbf{s}_k^{(i,j,t,\mathrm{op})} - \mathbf{s}_k^{(i,j,\iota)}$ and $\boldsymbol{\Sigma}_k^{(i,j,t,\iota)} = \boldsymbol{\Sigma}_k^{(i,j,t,\mathrm{op})} + \boldsymbol{\Sigma}_k^{(i,j,\iota)}$.

The Hungarian algorithm applied to $\mathbf{C}_{k,\mathrm{r},t,\iota}^{(i,j)}$ yields the optimal association, encoded as
\begin{equation}
    \mathbf{X}_{t,\iota}^{(i,j)} = \begin{cases}
        1 & \text{if tag } t \text{ matches hypothesis } \iota, \\
        0 & \text{otherwise}.
    \end{cases}
\end{equation}

\subsection{Communication Update Step}

Agent $a^{(i)}$ will have established a mapping between its own tags and the hypotheses transmitted by its neighbors. Since neighbors transmit only information about themselves and agent $i$, the received state is incomplete relative to $\mathbf{s}_k^{(i,\iota,t(\iota))}$ and $\mathbf{s}_k^{(i,t,\mathrm{op})}$. To handle this, agent $i$ constructs another set of extraction matrices $\mathbf{S}_{k,\mathrm{r}}^{(i,t,j)}$ and $\mathbf{S}_{k,\mathrm{r}}^{(i,\iota,t(\iota),j)}$, different from  $\mathbf{T}_{\mathrm{s}}^{(i,\iota,j)}$, and forms received information matrices and vectors: 
\begin{align}
    \begin{split}
        \mathcal{S}_k^{(\cdot)} = &\left\{\mathbf{S}_{k,\mathrm{r}}^{(\cdot)}\left(\boldsymbol{\Sigma}_{k,\mathrm{r}}^{(i,\nu,j)}\right)^{-1}\left(\mathbf{S}_{k,\mathrm{r}}^{(\cdot)}\right)^T \right. \\
         &\left.\mid \mathbf{X}_{t,\nu}^{(i,j)} = 1, j=1,\dots,\left|\mathcal{P}_k^{(i)}\right|\right\},\\
        \mathcal{V}_k^{(\cdot)} = &\left\{\mathbf{S}_{k,\mathrm{r}}^{(\cdot)}\left(\boldsymbol{\Sigma}_{k,\mathrm{r}}^{(i,\nu,j)}\right)^{-1}\boldsymbol{\mu}_{k,\mathrm{r}}^{(i,\nu,j)} \right. \\
         &\left.\mid \mathbf{X}_{t,\nu}^{(i,j)} = 1, j=1,\dots,\left|\mathcal{P}_k^{(i)}\right|\right\},\\
    \end{split}
\end{align}
where in the superscripts with $(\cdot)$ there is either $(i,t)$ or $(i,\iota,t(\iota))$ depending whether the operational hypotheses or tagged hypotheses are updated.

Covariance intersection is then applied as
\begin{equation}
    \begin{split}
        \left(\boldsymbol{\Sigma}_k^{(\cdot)}\right)^+ &= \left(\sum_{(j,\mathbf{I})\in\mathcal{S}_{k,\mathrm{r}}^{(\cdot)}}c_k^{(j,\cdot)}\mathbf{I} +c_k^{(i,\cdot)}\left(\boldsymbol{\Sigma}_k^{(\cdot)}\right)^{-1} \right)^{-1}, \\
        \left(\boldsymbol{\mu}_k^{(\cdot)}\right)^+ &= \left(\boldsymbol{\Sigma}_k^{(\cdot)}\right)^+\left(\sum_{(j,\mathbf{v})\in\mathcal{S}_{k,\mathrm{r}}^{(\cdot)}}c_k^{(j,\cdot)}\mathbf{v} +c_k^{(i,\cdot)}\mathbf{v}^{(\cdot)}\right). \\
    \end{split}
\end{equation}
where $\mathbf{v}^{(\cdot)} = \left(\boldsymbol{\Sigma}_k^{(\cdot)}\right)^{-1}\boldsymbol{\mu}^{(\cdot)}$. The weights satisfy $\sum_jc_k^{(j,\cdot)} + c_k^{(i,\cdot)}=1$. The superscripts with $(\cdot)$ is either $(i,t,\mathrm{op})$ or $(i,t,t(\iota))$ depending whether the operational hypotheses or the tagged hypotheses are updated. Likewise, the weights superscript is either given as $(i,t)$ or $(i,\iota,t(\iota))$.

The parameters $\left(\boldsymbol{\Sigma}_k^{(i,t,\mathrm{op})}\right)^+$, $\left(\boldsymbol{\Sigma}_k^{(i,\iota,t(\iota))}\right)^+$, $\left(\boldsymbol{\mu}_k^{(i,t,\mathrm{op})}\right)^+$ and $\left(\boldsymbol{\mu}_k^{(i,\iota,t(\iota)}\right)^+$ are treated as the priors in the next prediction step. 

\subsection{Tuning Covariance Intersection Weights}

The weights $c_k^{(j,t)}$, $c_k^{(i,t)}$, $c_k^{(j,\iota,t(\iota))}$, and $c_k^{(i,\iota,t(\iota))}$ are tuned to ensure rapid splitting of hypotheses in response to neighbors’ information, thereby enforcing timely attack detection across the network.  

Two special cases apply for the self-weight:  
\begin{itemize}
    \item If the agent maintains two or more tags with identical measurement support that include anchor measurements, then $c_k^{(i,t)} = c_k^{(i,\iota,t(\iota))} = 0.25$.  
    \item If no such duplicate tags exist then $c_k^{(i,t)} = c_k^{(i,\iota,t(\iota))} = 0.5$.  
\end{itemize}

Neighbor weights are further scaled relative to \ac{md}s computed in \eqref{eq:match_cost}, increasing separation between hypotheses and aligning them more strongly with neighbors. Specifically,  
\begin{equation}
    c_k^{(j,\cdot)} = \left(1-c^{(i,\cdot)}\right)\frac{\sum_{\nu}\mathbf{1}_{\mathbf{X}_{t,\nu}^{(i,j)}=1}\,d_{\mathrm{r}}^{(i,t,j,\nu)}}{\sum_{l\in\mathcal{P}_k^{(i)}}\sum_{\nu}\mathbf{1}_{\mathbf{X}_{t,\nu}^{(i,l)}=1}\,d_{\mathrm{r}}^{(i,t,l,\nu)}},
    \label{eq:CI-weight-average}
\end{equation}  
where for $(\cdot)$ in the superscripts one has $(t)$ or $(\iota,t(\iota))$.

\subsection{Increasing the Count of Hypotheses Tags}

The logic for incrementing hypothesis tags $t(\iota)$, introduced in \eqref{eq:agent_hypo}, is as follows.  

When neighbor $j$ detects an unlikely outlier count, it begins splitting hypotheses with varying subsets of measurement support. According to the rules in Section \ref{sec:transmitting}, $j$ then transmits a set of hypotheses $H_{k,\mathrm{r}}^{(i,j)}$ to agent $i$, see Section \ref{sec:utilizing}. This set may randomly contain more hypotheses than agent $i$ maintains. 

At initialization each agent is ready to increment the count of tags by one when at some $k$, it receives $|H_{k,\mathrm{r}}^{(i,j)}| > 1$. Once incremented the count of tags to $2$, the agents will only increment the number of tags, when they have seen $\tau_{\mathrm{n}}$ consecutive time steps where $|H_{k,\mathrm{r}}^{(i,j)}| > T_k^{(i)}$. There is a counter associated to each neighbour, such that when one of these counters exceed $\tau_{\mathrm{n}}$, the tag is increased. The unmatched hypothesis from agent $j$ in terms of the matching presented in Section \ref{sec:utilizing}, determines which existing tag will be cloned by looking at which tag is closest in terms of the cost matrix \eqref{eq:match_cost}. The operational hypothesis along with the hypotheses with the same tag will be copied and subsequently maintained in the following recursions.

\subsection{Initial Removal of Anchors as Measurement Support}

When an agent $a^{(i)}$ detects an abnormal count of outliers, and $T_k^{(i)} > 1$, and each tag only has one hypothesis, which occurs when recently the agent has increased the tag count, and it is within range of at least one anchor, following Assumption \ref{assum:sources}, the agents initial guess is to suspect the violating measurement support to be the anchor. Then, instead of splitting $|\mathcal{O}_k^{(i,\iota,t(\iota)}| - 1$ hypotheses, the anchor is removed from the measurement set of the affected hypothesis. This strongly separates the operational hypotheses in subsequent fusion steps.

\subsection{Computational Complexity}

The computational complexity of the \ac{ekf} prediction and update at any given time instance, assuming the \ac{ekf} has a complexity of $n_z^4$, where $n_x$ is the state dimension of $\mathbf{x}_k$, is given by $\sum_{\iota\in\mathcal{H}_k^{(i)}}\left(n_x + \left|\mathcal{O}^{(i,\iota)}\right|n_x\right)^4$. The Hungarian algorithm has a complexity of $n^3$, where $n$ is the dimension of the cost matrix in \eqref{eq:match_cost}. The Quickhull algorithm has a worst-case complexity of $n^2$, where $n$ is the number of points. Hence, the \ac{ekf} constitutes the worst scaling, especially since the number of hypotheses may increase rapidly.

For instance, consider an agent with a measurement support $\{0,1,2,3,4\}$, where the subsets $\{0,1,2\}$ and $\{3,4\}$ agree, respectively. The worst-case complexity occurs when the hypotheses split once, producing subsets including $\{0,1,2,3\}$ and $\{0,1,2,4\}$, such that the number of filters is five. Each filter is then estimating a state of dimension $5n_x$. This results in a total computational complexity scaling with the state dimension as $5(5n_x)^4$.

\section{Numerical Results}

This Section presents numerical results applied on a specific case study. In the case study each agent uses the following discrete motion model 
\begin{equation}
    \begin{split}
        \mathbf{x}_{k+1}^{(i)} &= \mathbf{x}_k^{(i)} + T_s
        \begin{bmatrix}
            \left(\mathbf{R}\left(\theta_k^{(i)}\right)\mathbf{v}_k^{(i)}\right)^\mathrm{T} & a_{x,k}^{(i)} & a_{y,k}^{(i)} & \omega_k^{(i)}
        \end{bmatrix}^\mathrm{T}, \\
        \mathbf{R}(\theta) &= 
        \begin{bmatrix} 
            \cos(\theta) & -\sin(\theta) \\ \sin(\theta) &  \cos(\theta)
        \end{bmatrix}, \\
        \mathbf{v}_k^{(i)} &= 
        \begin{bmatrix}
            v_{x,k}^{(i)} & v_{y,k}^{(i)}
        \end{bmatrix}^{\mathrm{T}},
    \end{split}
    \label{eq:scenario_motion_model}
\end{equation}
where $v_{x,k}$, $a_{x,k}$, $v_{y,k}$ and $a_{y,k}$ are the velocities and accelerations in $x$ and $y$ directions respectively, and $\omega$ is the angular velocities. $a_{x,k}$, $a_{y,k}$ and $\omega$ are the inputs to the systems and are measurements from an \ac{imu}. $T_s$ is the sampling period. The process noise covariance matrices are $\mathbf{Q}^{(i)} =\mathbf{Q}^{(i,j)} =\texttt{diag}\left(\begin{bmatrix} \sigma_x^2 & \sigma_y^2 & \sigma_{v_x}^2 & \sigma_{v_y}^2 & \sigma_{\theta}^2 \end{bmatrix}\right)$ with $\sigma_x^2 = \sigma_y^2 = 0.5\mathrm{m}^2$, $\sigma_{v_x}^2 = \sigma_{v_y}^2 = 10^{-4}\frac{\mathrm{m}^2}{\mathrm{s}^2}$ and $\sigma_\theta = \frac{\pi}{1800}\mathrm{rad}^2$, $\mathbf{R}_{\mathrm{IMU}} = \texttt{diag}\left( \begin{bmatrix} \sigma_{a_x}^2 & \sigma_{a_x}^2 &  \sigma_\omega^2 \end{bmatrix}\right)$ where $\sigma_{a_x}^2 = \sigma_{a_y}^2 = 10^{-3}\frac{\mathrm{m}^2}{\mathrm{s}^4}$ and $\sigma_\omega = 2\cdot10^{-5}\frac{\mathrm{rad}^2}{\mathrm{s}^2}$, and $\mathbf{R}_\mathrm{RF} = \texttt{diag}\left(\begin{bmatrix} \sigma_r^2 & \sigma_\mathrm{AOA}^2 & \sigma_\mathrm{AOD}^2 \end{bmatrix}\right)$, where $\sigma_r^2 = 1\mathrm{m}^2$, $\sigma_\mathrm{AOA}^2 = \sigma_\mathrm{AOD}^2 = \frac{\pi}{360}\mathrm{rad}^2$. The agents use the same model in \eqref{eq:scenario_motion_model} to track their neighbors. 

\begin{figure}[!t]
    \centering
    \includegraphics[width=0.8\linewidth]{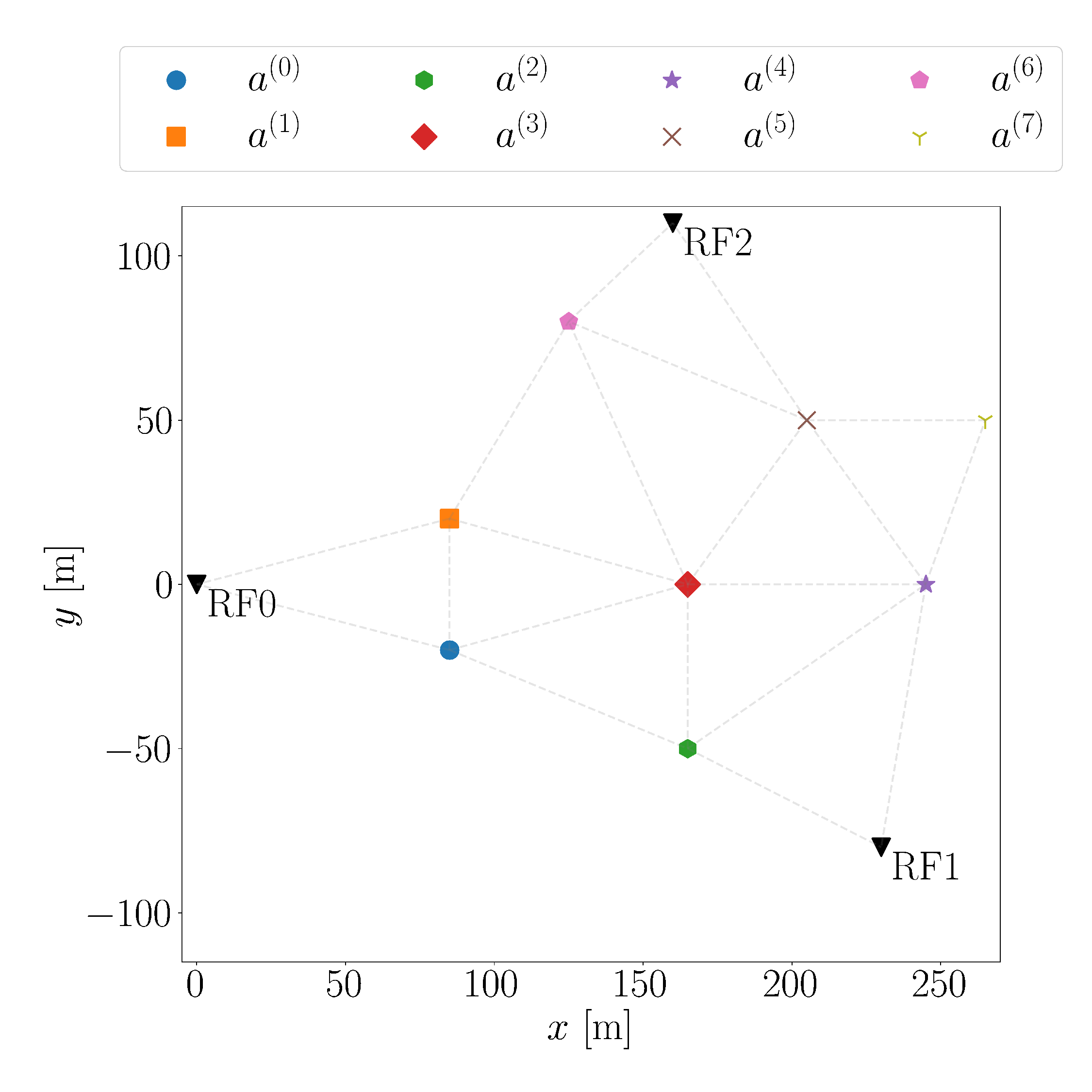}
    \caption{Multi-Agent system}
    \label{fig:scenario}
\end{figure}

\begin{figure}[!t]
    \centering
    \footnotesize
    \subfloat[Agent $a^{(0)}$ at $k=55$]{
        \includegraphics[width=0.48\linewidth]{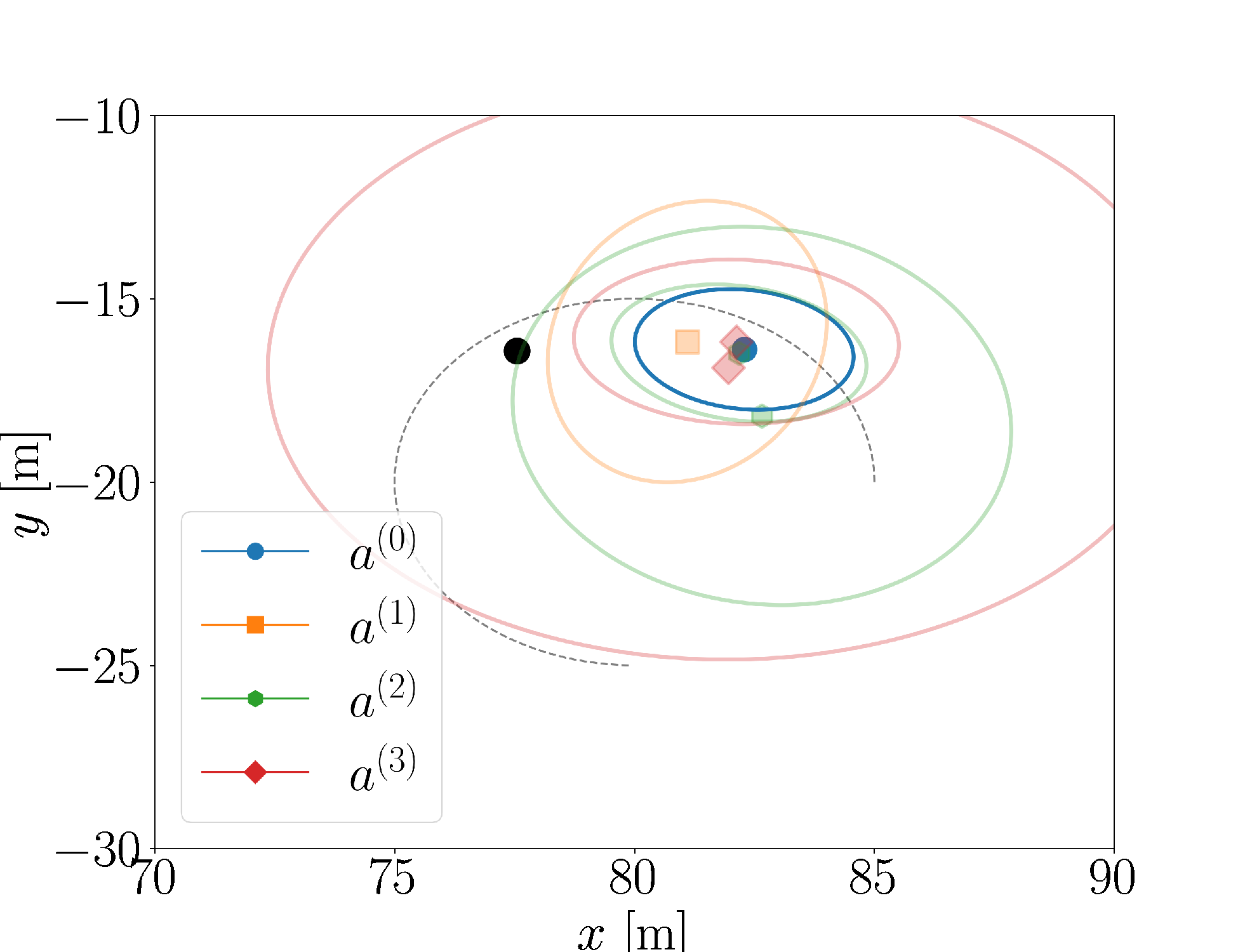}
    }
    \subfloat[Agent $a^{(7)}$ at $k=55$]{
        \includegraphics[width=0.48\linewidth]{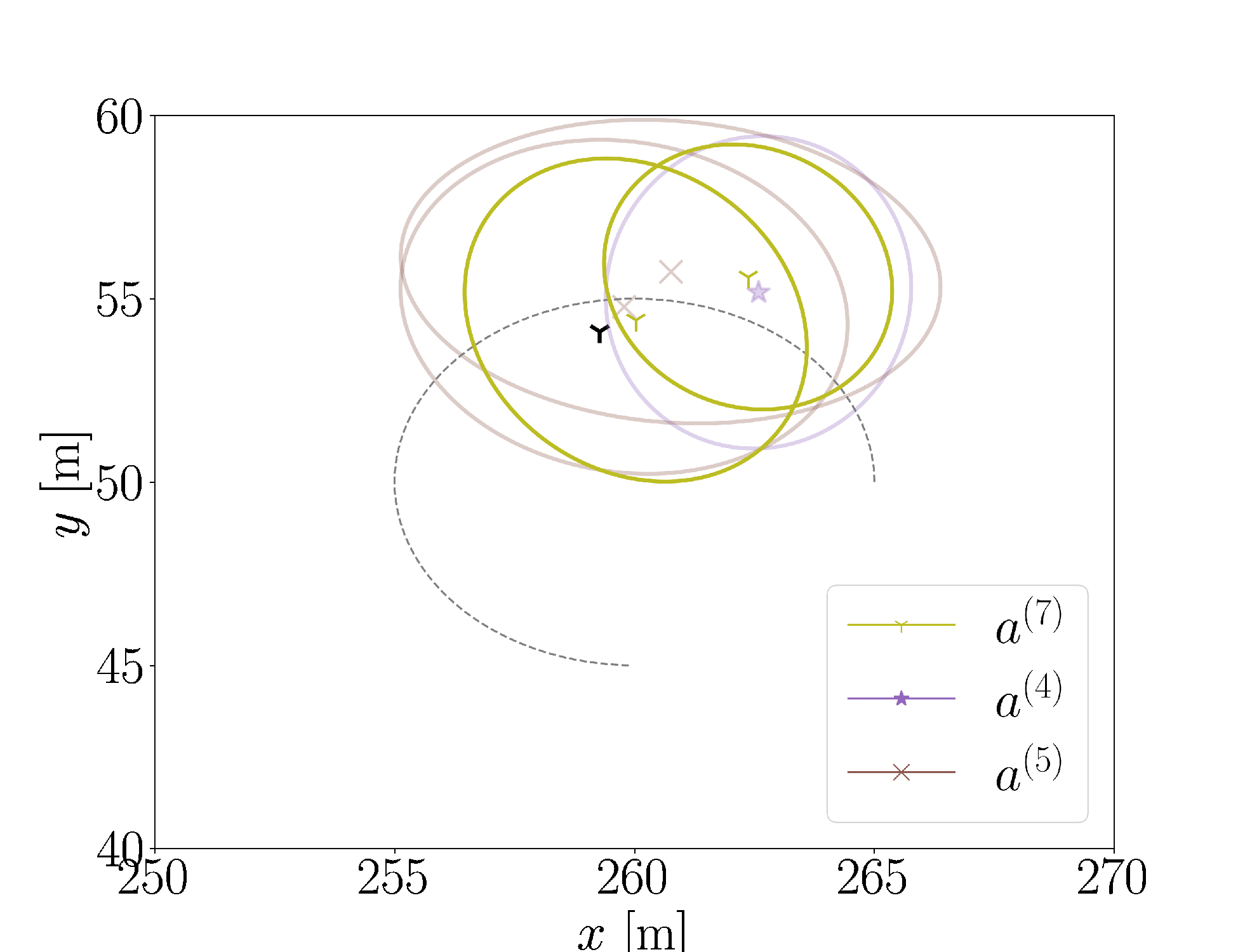}
    } \\
    \subfloat[Agent $a^{(0)}$ at $k=57$]{
        \includegraphics[width=0.48\linewidth]{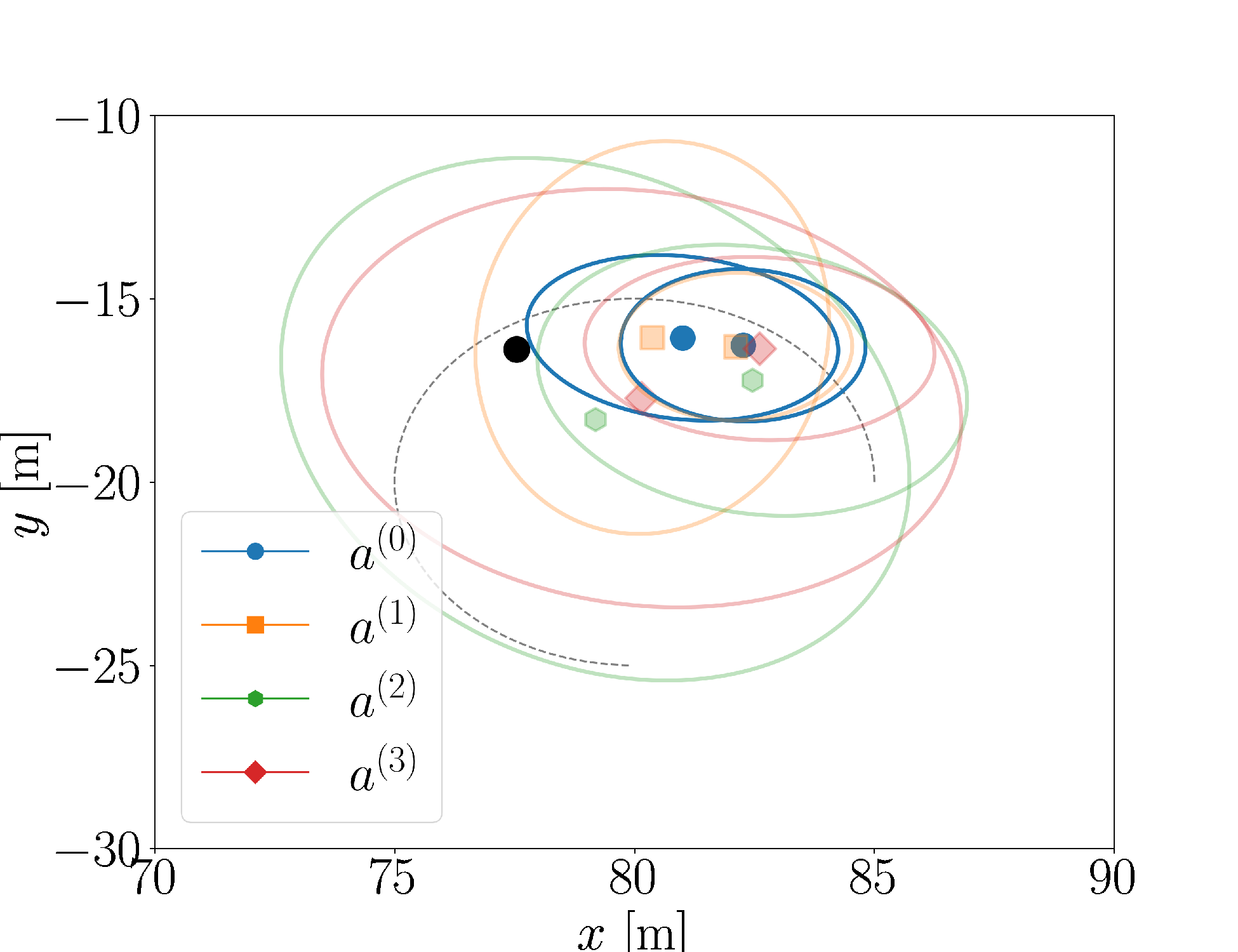}
    }
    \subfloat[Agent $a^{(7)}$ at $k=57$]{
        \includegraphics[width=0.48\linewidth]{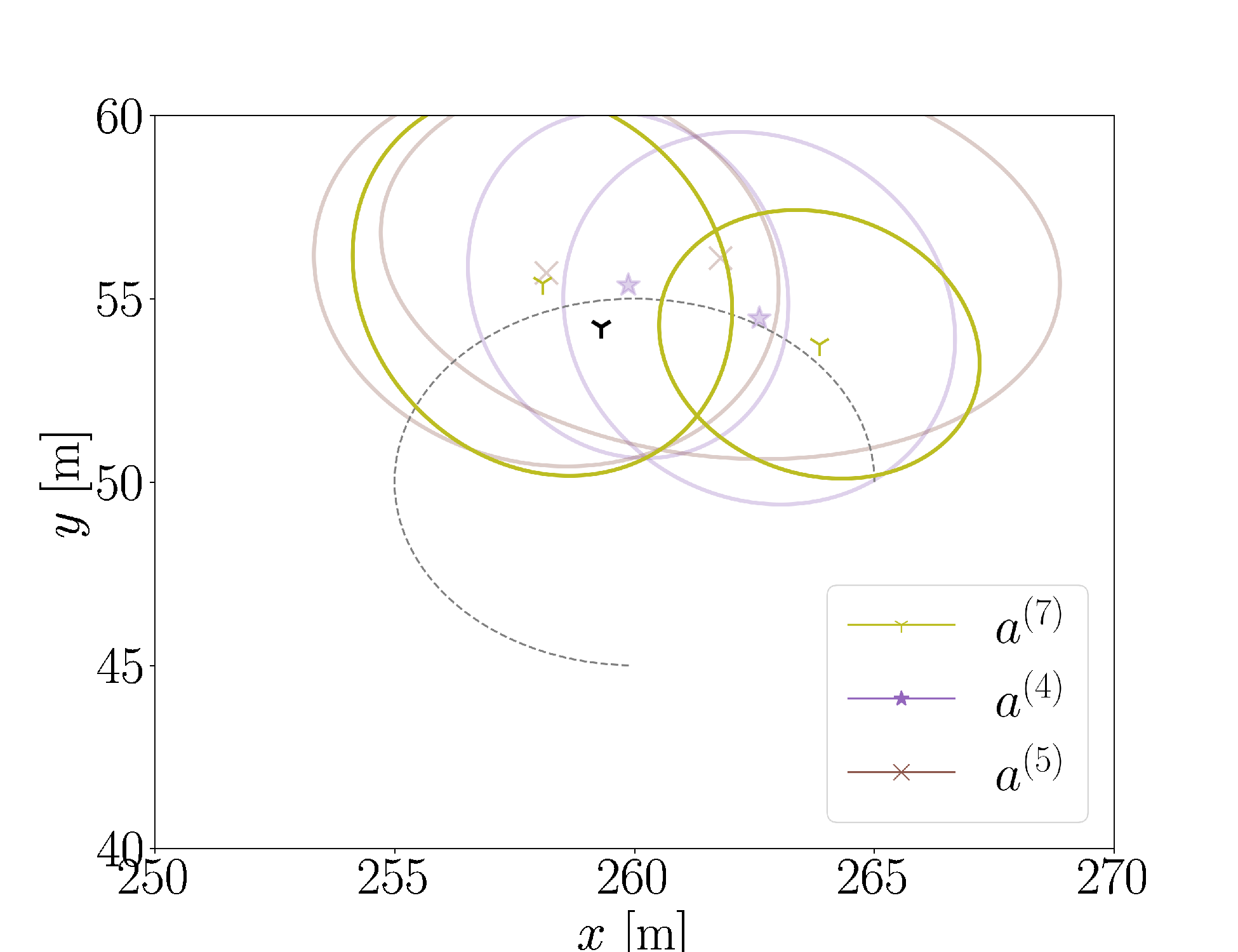}
    }\\
    \subfloat[Agent $a^{(0)}$ at $k=165$]{
        \includegraphics[width=0.48\linewidth]{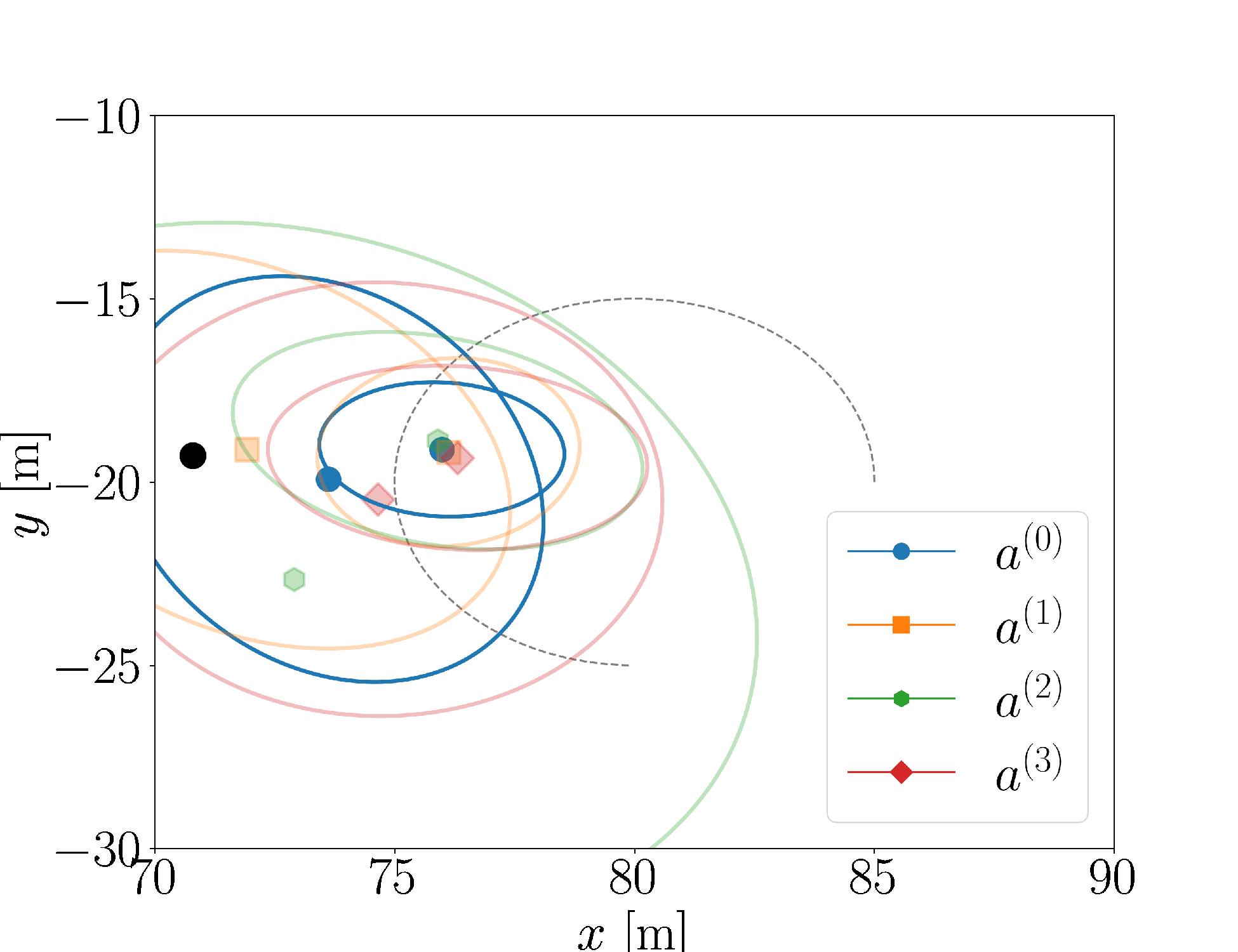}
    }
    \subfloat[Agent $a^{(7)}$ at $k=165$]{
        \includegraphics[width=0.48\linewidth]{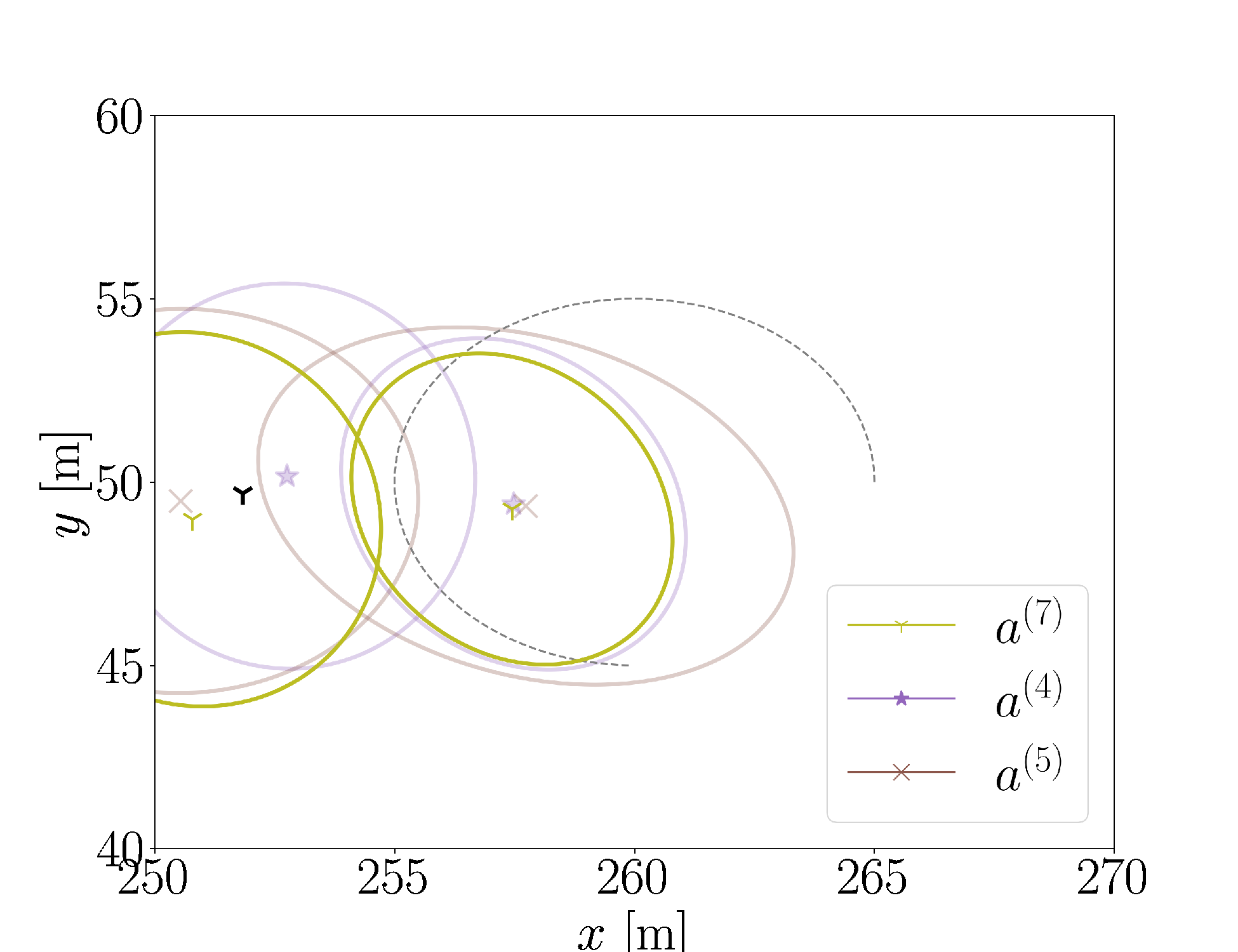}
    }\\
    \subfloat[Agent $a^{(0)}$ at $k=222$]{
        \includegraphics[width=0.48\linewidth]{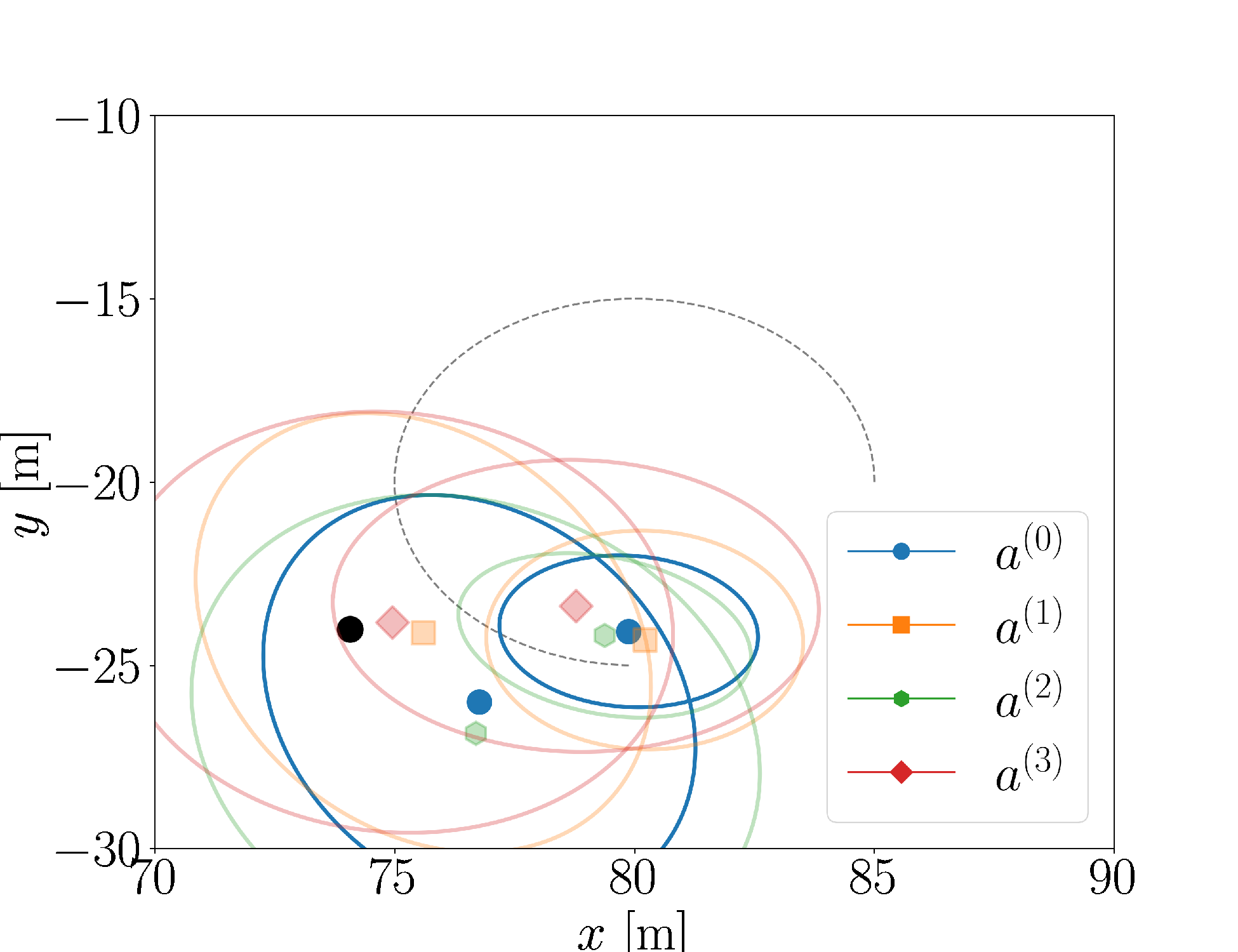}
    }
    \subfloat[Agent $a^{(7)}$ at $k=222$]{
        \includegraphics[width=0.48\linewidth]{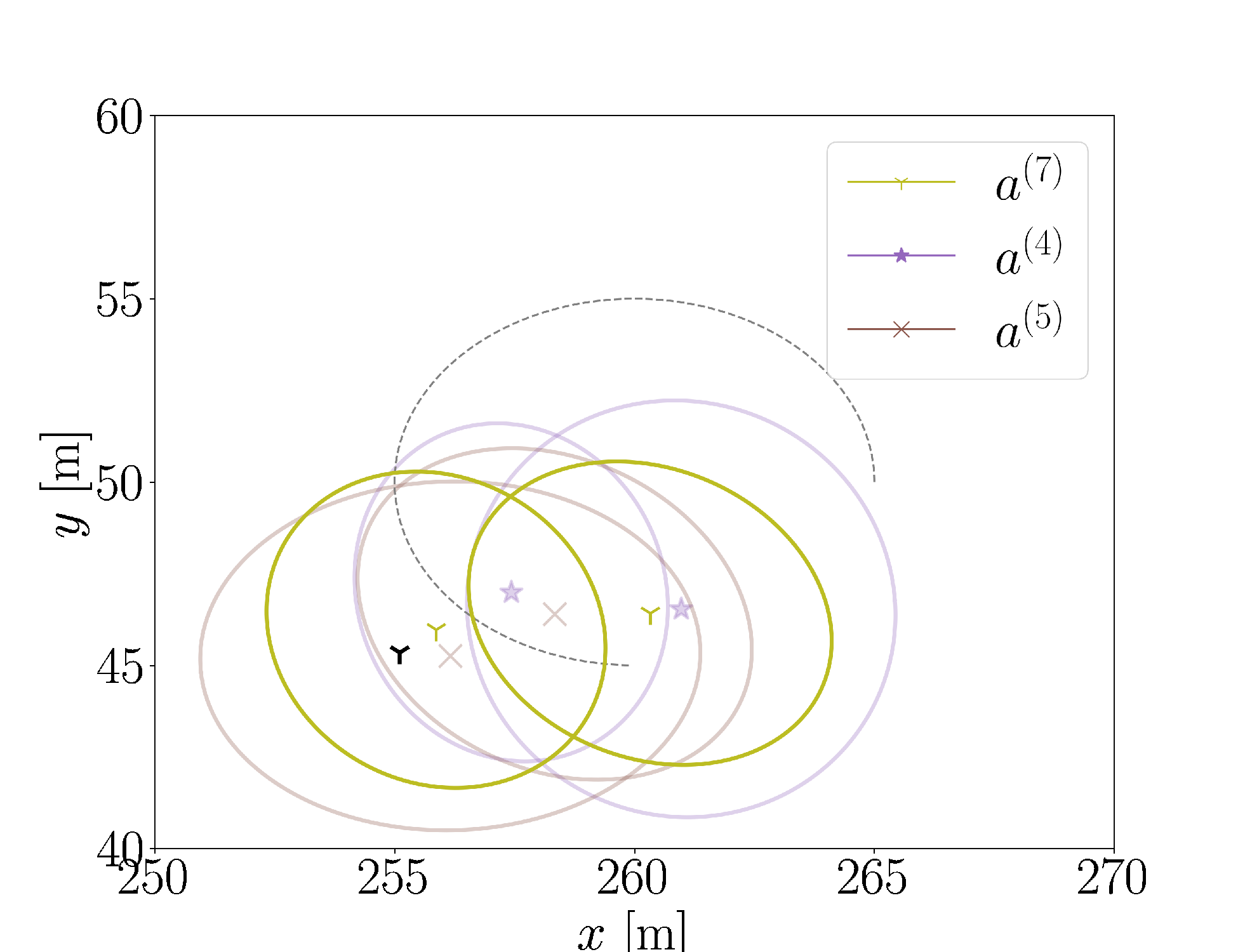}
    }
    \caption{Specific time steps showing the operational hypotheses $h^{(i,t,\mathrm{op})}$ of agents $a^{(0)}$ and $a^{(7)}$ along with the neighbors estimate of their position The plots show the error ellipses with containment probability $99.73\%$. The relative positions of agents are shown in Fig. \ref{fig:scenario}. The attack occurs at time step $k=20$, affecting the measurements relative to anchors $\mathrm{RF}0$ and $\mathrm{RF}1$. At time step $k=55$ agent $a^{(7)}$ has received more than one hypothesis from one of its neighbour, adding a new tag such that new operational hypothesis is added. The same occurs at at agent $a^{(0)}$ at $k=57$. Time steps $k=165$ and $k=222$ shows how the true state fall within the error ellipses. The black dot represents the true coordinates of the agents. The gray curve shows the nominal trajectory, which the agent is supposed to track.}
    \label{fig:results}
\end{figure}

The case study considers eight agents and three anchors, shown in Fig. \ref{fig:scenario}. A malicious attacker spoofs the anchors $\mathrm{RF}0$ and $\mathrm{RF}1$ such that the agents are biased $5\mathrm{m}$ in the $x$-coordinate. Each agent does a circular maneuver, where the circle has a radius of five meters. Fig. \ref{fig:results} show the operational hypotheses at specific time steps for agents $a^{(0)}$ and $a^{(7)}$. These agents are shown since $a^{(0)}$ together $a^{(2)}$ are those furthest from the truthful $\mathrm{RF}2$, and $a^{(7)}$ is the furthest from $\mathrm{RF}1$ and $\mathrm{RF}2$. 

A metric used in collaborative localization is the \ac{nees}. It measures the property of the estimator to be consistent. Since this paper introduces a change in the typical \ac{ci} fusion protocol in that only the agents may share their own information together with the neighbours information, the \ac{nees} is shown in four different cases; (1) where there is no attack and the agents only transmit their own information together with the recipients information, (2) there is no attack and the agents share all information with their neighbours, (3) there is an attack and information is shared as in (1), and (4) there is an attack and the information is shared as in (2). 

The \ac{nees} is presented only based on the pose of the agents. Denote $\boldsymbol{\mu}_{k,\mathrm{q}}^{(i)}$ and  $\mathbf{P}_{k,\mathrm{q}}^{(i)}$ as the mean value and covariance matrix agent $a^{(i)}$'s pose. The \ac{nees} is hence computed as 
\begin{equation}
    d_{\mathrm{NEES}} = \left(\mathbf{e}_{k,\mathbf{q}}^{(i)}\right)^T\left(\mathbf{P}_{k,\mathrm{q}}^{(i)}\right)^{-1}\mathbf{e}_{k,\mathbf{q}}^{(i)},
\end{equation}
where $\mathbf{e}_{k,\mathbf{q}}^{(i)} = \mathbf{q}_k^{(i)} - \boldsymbol{\mu}_{k,\mathrm{q}}^{(i)}$.

\begin{figure}[!t]
    \centering
    \subfloat[Agent $a^{(0)}$]{
        \includegraphics[width=0.48\linewidth]{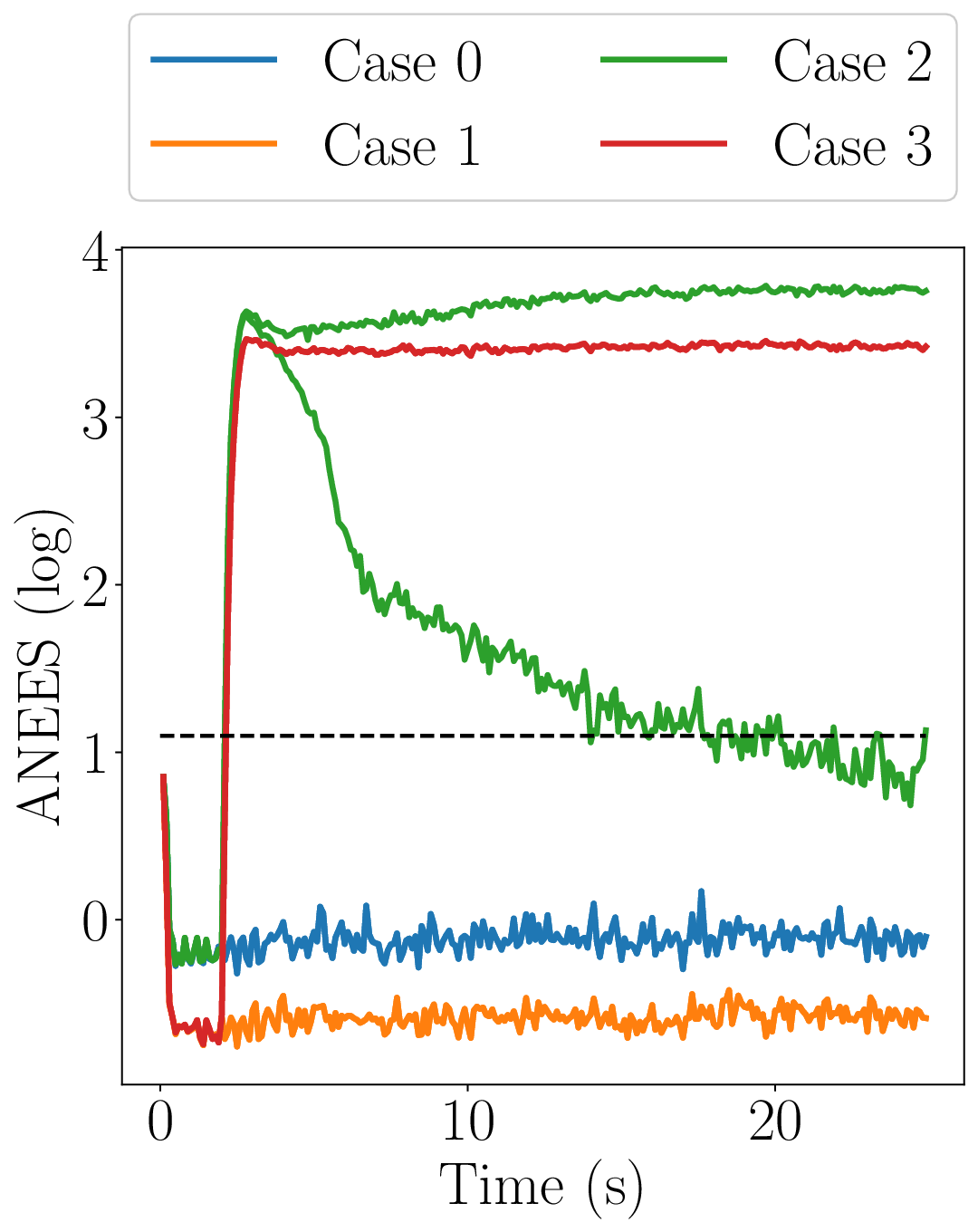}
    }
    \subfloat[Agent $a^{(7)}$]{
        \includegraphics[width=0.48\linewidth]{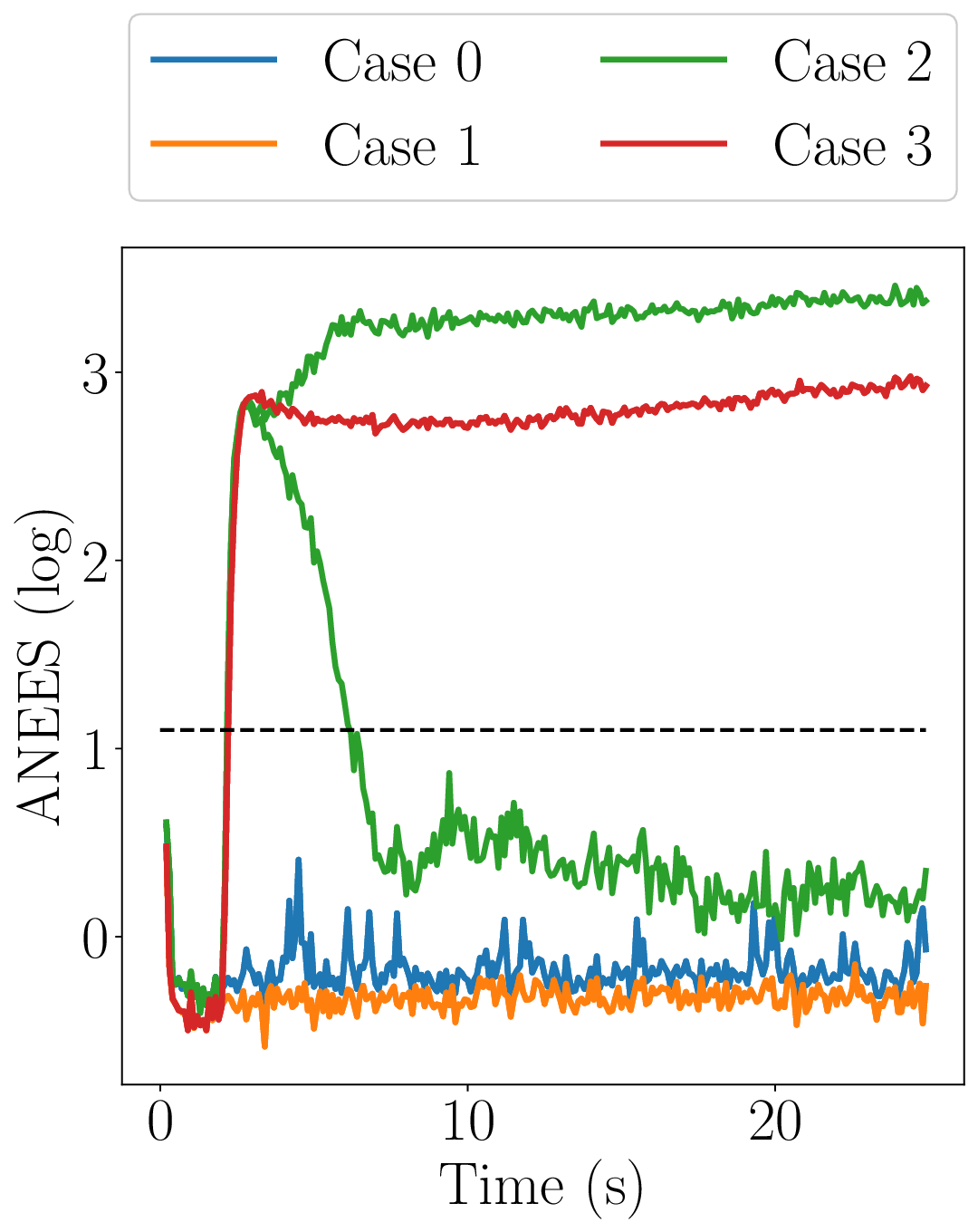}
    } 
    \caption{The average \ac{nees} over 200 noise realizations and using the scenario shown in Fig. \ref{fig:results}. The cases are: (1) where there is no attack and the agents only transmit their own information together with the recipients information, (2) there is no attack and the agents share all information with their neighbours, (3) there is an attack and information is shared as in (1) and (4) there is an attack and the information is shared as in (2). The two curves for case 2 show that the algorithm creates another tag within on average 40 samples. The black dashed line shows the expected value of the \ac{nees}, $\mathbb{E}(d_\mathrm{NEES})=3$.}
    \label{fig:results_ANEES}
\end{figure}

The average \ac{nees} results are shown in Fig. \ref{fig:results_ANEES}. The figure shows a dashed black line which represents the expected value of the \ac{nees}. A filter which is consistent is supposed to be equal to this line as the number of realization increases.  Anything below that line is said to conservative and anything above is inconsistent or overconfident. In the cases where there is no attack, the filter is conservative which is expected of the \ac{ci} collaborative localization method. The method proposed in this paper creates two tags when an attacker is present, shown in green. It can be seen, that one of the tags is consistent, representing the hypotheses tracking the true state, and a second, tracking the spoofed state. This is visualized in Fig. \ref{fig:results} for specific time steps. The figure shows how prior to the splitting of hypotheses the agent is overconfident. Right when the split occurs, it remains overconfident, but regains its consistency as time progresses. Agent 0's consistent hypothesis has a relatively large covariance matrix, which is due to to being the most distant to the non-adversarial anchor.

\begin{figure}[!t]
    \centering
    \subfloat[Agent 0]{
        \includegraphics[width=0.48\linewidth]{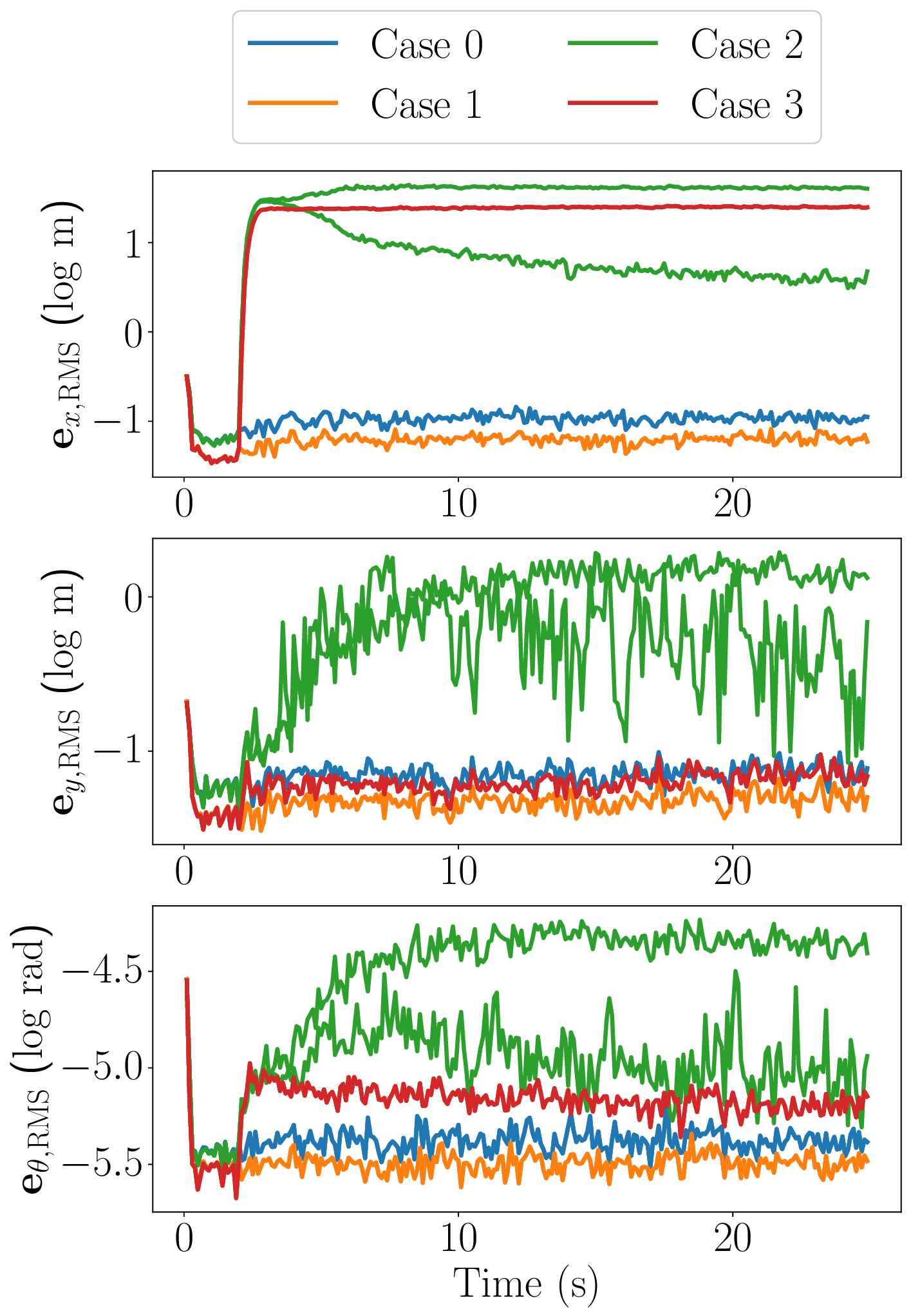}
    }
    \subfloat[Agent 7]{
        \includegraphics[width=0.48\linewidth]{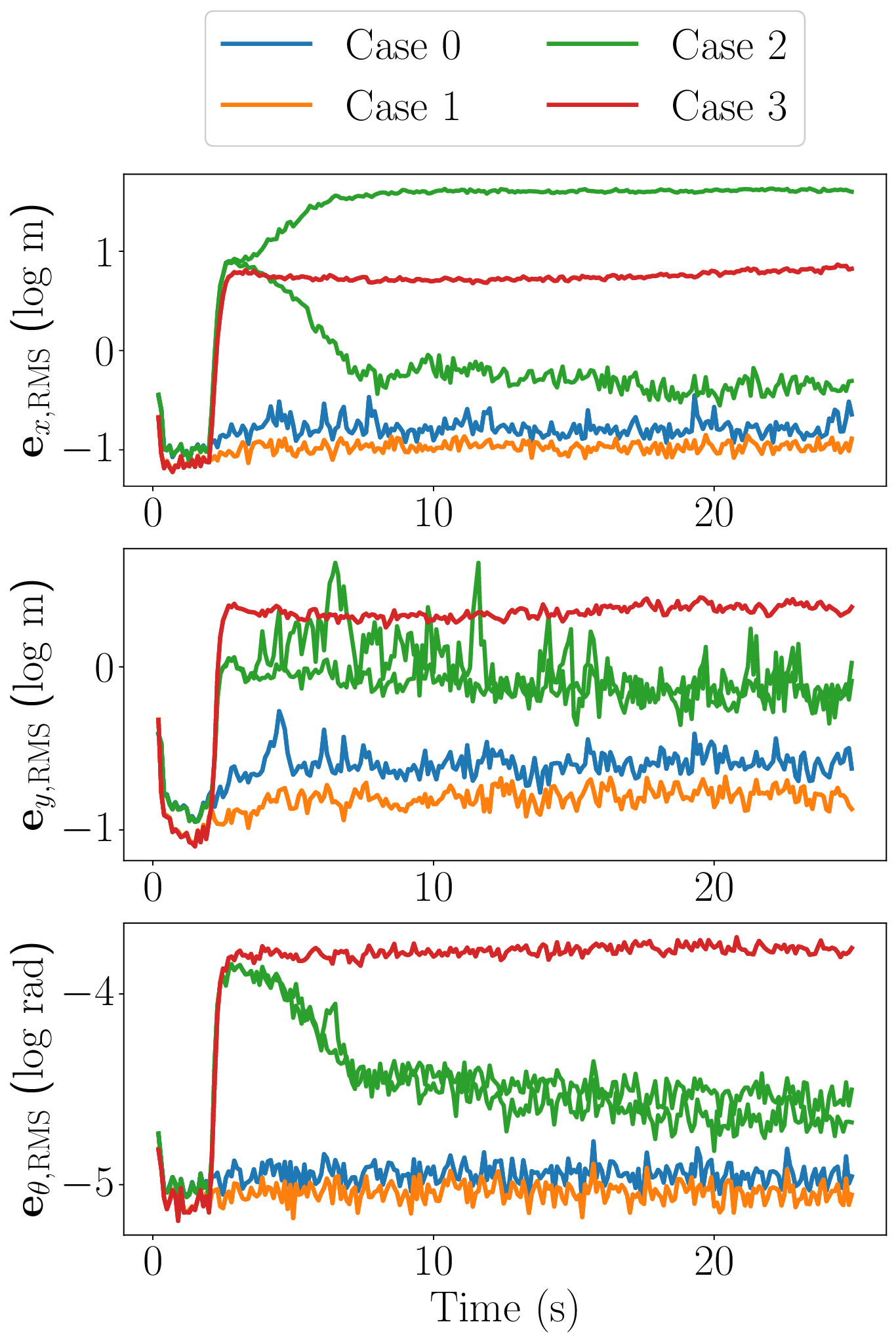}
    } 
    \caption{The \ac{rms} of the $x$- and $y$-coordinates and the heading $\theta$, computed using 200 noise realizations.}
    \label{fig:results_RMS}
\end{figure}

The \ac{rms} of the pose is shown in Fig. \ref{fig:results_RMS}. It can be seen, that the attack, although only effectuated in the $x$-coordinates, also affects the $y$-coordinate and the heading. Is can be seen, that the algorithm does not regain the same navigational performance. The hypotheses tracking the true state have effectively lost a measurement source, and will naturally have lower accuracy. It can be seen, that the loss in navigation performance is lower when using the proposed algorithm compared to the naive fusion, where no attack is anticipated.

\section{Discussion}

The method presented in this paper is based on \ac{ci} based collaborative localization. \ac{ci} is known to be a conservative fusion method, since it assumes maximum correlation, with the consequence of inflating the covariance matrix. This impacts the windowed count detector in the sense that the outlier probability increases, which in turns implies that the count of outliers over the window $W$ needs to be relatively high. The numerical results provided in the previous Section use a bias magnitude of five meters, which is relatively large compared to the noise parameters. As each hypothesis effectively needs to see a high count of outliers, this delays the detection and conclusion of the diagnosis at the individual agents and for the system as shown in Fig. \ref{fig:results_ANEES} and \ref{fig:results_RMS}. 

The complexity of the algorithm grows greatly as the number of measurement sources increases, which is also true for the \ac{ci} collaborative localization method used in \cite{chang_resilient_2022}. As was shown in \cite{roumeliotis_analysis_2003}, adding more agents has a diminishing return in increasing the accuracy of the system. Following this, one would have to plan the nominal trajectories $\mathbf{t}_\mathrm{n}^{(i)}$ such that collaborative localization is done with five neighbors, which is where the gain in accuracy drops relatively \cite{roumeliotis_analysis_2003}.

Instead of using the convex hull approach to determine the number of hypotheses to transmit, it would seem appropriate to use a clustering method, such as $k$-means clustering or spectral clustering. In both clustering methods, a number of $k$ clusters should be selected apriori and is typically determined based on data or expert knowledge. In spectral-clustering one could use the eigengap heuristic \cite{von_luxburg_tutorial_2007} to determine the number of clusters. However, it was found that the distances in $\mathbf{D}_{k,(\iota,\nu)}^{(i,j)}$ would lead to an inconsistent number of clusters and could suddenly jump to a large number of clusters. 

The \ac{ci} has an averaging effect. To enforce the seperation of the hypotheses, \eqref{eq:CI-weight-average} scales the weights according to the \ac{md} evaluation, with the intention to separate the hypotheses. In a constellation where agent $a^{(i)}$ is a neighbor to agents $a^{(j)}$ and $a^{(l)}$, who are not neighbors of each other, agent $a^{(j)}$ will have a relatively high \ac{md} evaluation, moving $a^{(i)}$s  estimate towards agent $a^{(j)}$s estimate. In a subsequent fusion step however, agent $a^{(l)}$ will now have a relatively high \ac{md} evaluation such that it brings agents $a^{(i)}$s estimate back. This will prolong the separation of the hypotheses, therefore also the ability of the system to propagate the hypotheses and in turn also the ability of each individual agents to detect the spoofing attack.

\section{Conclusion}

This paper presented a method for resilient collaborative localization in multi-agent systems subject to attacks on \ac{rf} measurements. The approach extends a multi-hypotheses framework to the networked setting by introducing tagged hypotheses and fusion through \ac{ci}. A set of reductions based on distance tests and convex hull geometry limits the number of hypotheses that agents share, which enables the network to propagate diagnostic information without excessive communication.

The results show that the method allows agents to identify and separate spoofed measurements and to recover estimates that remain consistent with the true state once the correct hypothesis is isolated. The reduction in accuracy that follows from removing compromised measurements and the conservative nature of covariance intersection is expected, and the study also indicates that detection slows due to relatively large covariance matrices. The approach can be strengthened by reducing conservativeness in the fusion process and improving hypothesis management. A next step is to investigate the coordination between agents to accept a single hypothesis across the whole network.

\small
\bibliographystyle{IEEEtran}
\bibliography{references.bib}

\newpage

 




\vfill

\end{document}